Министерство образования и науки Украины
Харьковский национальный университет имени В. Н. Каразина

Н. Т. Гладких, С. В. Дукаров, В. И. Фареник

# СМАЧИВАНИЕ
# В ОСТРОВКОВЫХ ПЛЕНКАХ

Учебное пособие
к спецкурсу «Физика тонких пленок»

Харьков – 2003

Учебное пособие к спецкурсу «Физика тонких пленок». В пособии изложены современные методы исследования смачивания в нанодисперсных системах, основанные на применении электронной микроскопии, и экспериментальные результаты о проявлении размерных эффектов при контактном взаимодействии высокодисперсных фаз.

Учебное пособие предназначено для студентов физических и физико-технических специальностей высших учебных заведений.

Рецензенты: заведующий кафедрой металлофизики Национального технического университета (Харьковский политехнический институт), доктор физ.-мат. наук, проф. А. Т. Пугачев;
заслуженный профессор Харьковского национального университета им. В. Н. Каразина, доктор физ.-мат. наук, проф. А. Ф. Сиренко.

# Введение

Проблема смачивания в конденсированных пленках непосредственно связана с самим процессом их формирования и с общей проблемой образования жидкой фазы при конденсации пересыщенного пара на твердых поверхностях. Это обусловлено тем, что в существующих представлениях о конденсации пересыщенного пара, основанных на зародышевом механизме, в качестве определяющего параметра, наряду с поверхностной энергией, является краевой угол смачивания жидкой фазой твердой поверхности. Однако, несмотря на это, специальные исследования смачивания в островковых конденсатах не проводились, а при необходимости использовались имеющиеся данные о краевых углах смачивания для макроскопических систем, полученные с применением традиционных методов.

Необходимость исследования смачивания непосредственно на островковых конденсатах возникла в связи с потребностью в таких данных при выяснении физической природы образования и устойчивости жидкой фазы в островковых пленках ниже температуры плавления. При этом обнаружилось, что применение традиционных методов исследования смачивания для таких объектов ограничено или вообще невозможно. Поэтому одновременно с изучением переохлаждения при кристаллизации в островковых пленках совершенствовались существующие методы и разрабатывались новые. Параллельно с развитием новых методов, особенно основанных на использовании электронной микроскопии, появились экспериментальные возмож-ности изучения более тонких эффектов, уже связанных с размерной зависимостью самой поверхностной энергии малых частиц и тонких пленок.

В то же время расширение исследований физико-химических свойств малых частиц и их ансамблей, а также различные применения нанодисперсных систем в свою очередь стимулировали изучение поверхностных явлений в островковых конденсатах, которые являются удобными объектами для моделирования таких систем.

Исходя из этого, наряду с обсуждением результатов по исследованию смачивания в конденсированных пленках, рассматриваются и разработанные для этих целей экспериментальные методы. Представляется, что приведенные результаты являются важными не только для физики островковых и сплошных пленок, но имеют и самостоятельное значение для физики и физико-химии поверхности и поверхностных явлений, особенно для нанодисперсных систем.

*Глава 1*

## Методы определения смачивания в высокодисперсных системах

Важнейшим параметром, определяющим капиллярные свойства систем, является величина соответствующей поверхностной энергии (собственная поверхностная энергия для изолированных объектов либо межфазная энергия контактирующих фаз); сведения об этих величинах могут быть получены при исследовании смачивания в двухфазных системах типа твердое тело – жидкость. Анализ известных возможностей определения краевых углов смачивания $\theta$ показывает, что традиционные методы, подробный обзор которых приведен в работах [1–5], весьма ограничено применимы для изучения смачивания в ультрадисперсных системах, поскольку большинство из них требуют значительных количеств жидкости, а метод капли на тонкой нити ограничен случаем $\theta < 90°$. В связи с этим очевидна необходимость разработки новых методик, которые позволили бы исследовать смачивание и контактное взаимодействие в ультрадисперсных системах, причем как для систем с дисперсной жидкой фазой, так и для систем с дисперсной твердой фазой. Учитывая, что для ультрадисперсных систем следует ожидать проявления размерных эффектов, методики должны быть применимы в широком диапазоне размеров.

В работах [6–12] приведены методики, позволяющие исследовать смачивание в системах с различным типом контакт-ного взаимодействия (т. е. применимые как при $\theta < 90°$, так и для $\theta > 90°$) при изменении характерных размеров фаз в пределах $3–10^5$ нм.

Образцы для исследований представляли собой островковые пленки различных металлов, конденсированные в вакууме по механизму пар→жидкость на твердые подложки, которые, как правило, также препарировали методом вакуумной конденсации.

Использование метода вакуумной конденсации для препарирования образцов позволило обеспечить высокую чистоту контактирующих поверхностей и минимизировать влияние на смачивание факторов физико-химического происхождения, таких как образование адсорбционных слоев, оксидных пленок и т. д. Метод вакуумной конденсации дает возможность устранить влияние геометрических факторов (шероховатость и пористость подложки и т. д.), поскольку поверхность последней может быть гладкой и «чистой» на атомарном уровне, в случае применения вакуумных сколов и свеженапыленных в условиях сверхвысокого вакуума слоев. И, наконец, метод вакуумной конденсации позволяет контролируемым образом варьировать степень дисперсности системы капля – подложка и, таким образом, исследовать влияние размерного фактора на смачивание.

Для получения образцов [8, 9, 11] использовались вакуумные установки с безмасляной системой откачки, позволяющие препарировать пленки в вакууме $10^{-7}–10^{-9}$ мм рт. ст. Методика препарирования образцов была следующей. На скол монокристалла NaCl (или KCl) в вакууме при температуре подложки, обеспечивающей конденсацию исследуемого металла в жидкую фазу, наносили слой вещества, выбранного в качестве подложки, а затем по механизму пар–жидкость конденсировался исследуемый металл. Массовую толщину пленок определяли с помощью кварцевого датчика, а также по контрольным стеклам методом линий равного хроматического порядка или по оптической плотности образца. Полученные пленки охлаждали в вакууме до комнатной температуры и далее закристаллизовавшиеся частицы исследовали методами оптической, растровой и просвечивающей электронной микроскопии.

В случае, когда можно пренебречь влиянием силы тяжести, форма малых капель представляет собой сферический сегмент (оценки показывают, что для металлов это заведомо справедливо при размерах частиц менее $10^5$ нм). Для нахождения угла $\theta$ достаточно измерить любые две из трех величин, характеризующих капли на подложке: радиус кривизны поверхности капли $R$, диаметр ее основания $d$ ($d = 2r$) и высоту $H$ (рис. 1а).

В изложенных в работах [7–12] методах указанные величины измеряли в основном на закристаллизовавшихся каплях, поэтому была оценена погрешность [12], обусловленная изменением объема капли при ее затвердевании на подложке. Поскольку теплоотвод при охлаждении капель в вакууме осуществляется в основном через границу контакта капли с подложкой, то при кристаллизации капель площадь контакта не изменяется. В этом случае, как было показано [12], погрешность в определении $\theta$ равна

$$\Delta\theta = \frac{1}{3}(2 + \cos\theta)\sin\theta\frac{\Delta V}{V_s} \ , \tag{1}$$

где $\Delta V/V_s$ – относительное изменение объема при кристаллизации.

Оценки [12] показывают, что при $\Delta\rho/\rho_s = 0{,}05$ (характерная величина для типичных металлов) $\Delta\theta$ не превышает 2°. Это подтверждается также экспериментальными данными [7, 8], согласно которым при кристаллизации частиц краевой угол в пределах погрешностей измерений не изменяется. На рис. 2 приведены микроснимки профилей металлических капель, полученные до и после их плавления электронным пучком. Видно, что изменение краевого угла при фазовом переходе незначительно и находится в пределах приведенных оценок.

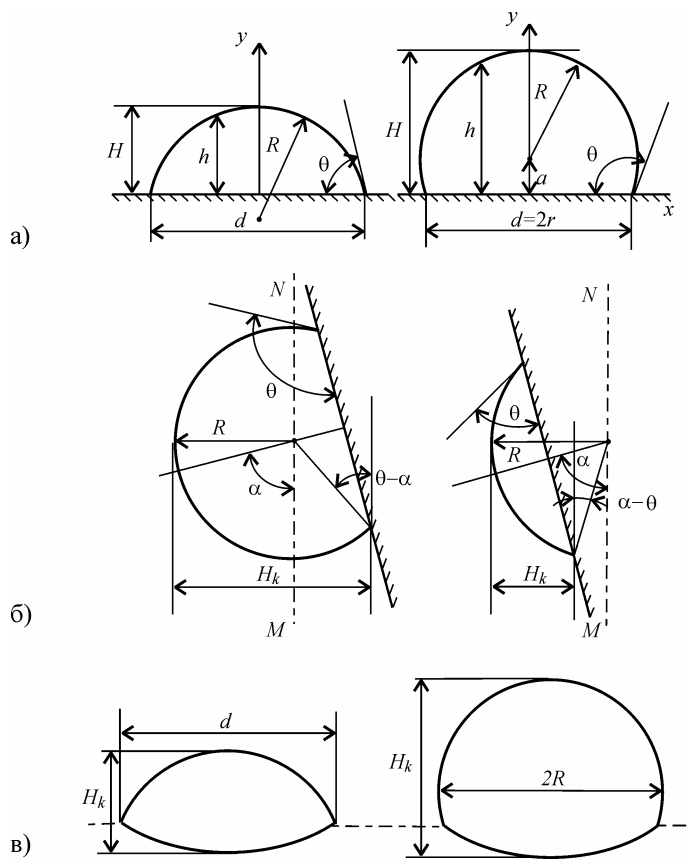

*Рис. 1. Схематическое изображение (а) жидкой капли на твердой подложке, сечение (б) и вид (в) капель при наклонном наблюдении образца [12]*

Таким образом, изменением угла при кристаллизации жидких капель можно пренебречь и относить величины θ, найденные для закристаллизовавшихся частиц, к величинам краевых углов жидких капель при температуре их образования.

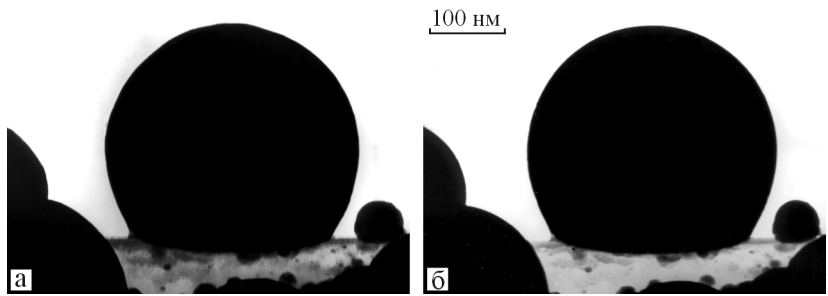

*Рис. 2. Электронно-микроскопические снимки кристаллических (а) и жидких (б) частиц свинца на углеродной подложке*

Для измерения углов смачивания подложек микрокаплями размером $10^3$–$10^5$ нм используются методы оптической микроскопии.

**Метод скола** [7, 8] основан на измерении параметров профиля капли в плоскости, перпендикулярной поверхности подложки (плоскость скола подложки (рис. 1а и 3а)). В этом случае легко могут быть измерены все три параметра $R$, $d$ и $H$, а угол $\theta$ определяется соотношением

$$\theta = 2\,\text{arctg}\,\frac{2H}{d} = \arccos\left(1 - \frac{H}{R}\right) = \begin{cases} \arcsin\dfrac{d}{2R}, \theta < 90°, \\ 180° - \arcsin\dfrac{d}{2R}, \theta > 90°. \end{cases} \quad (2)$$

Так как при вакуумной конденсации характерно распределение частиц по размерам, то в плоскости скола всегда оказывается достаточное для измерений количество частиц существенно различных размеров. В этом методе для получения надежных значений угла $\theta$ величины $R$, $d$ и $H$ следует по возможности измерять для частиц, находящихся точно в плоскости скола. Поскольку в оптической микроскопии глубина резкости невелика, то критерием правильности выбора частиц для измерений углов $\theta$ является одновременно резкое изображение основания частицы и края скола кристалла-подложки.

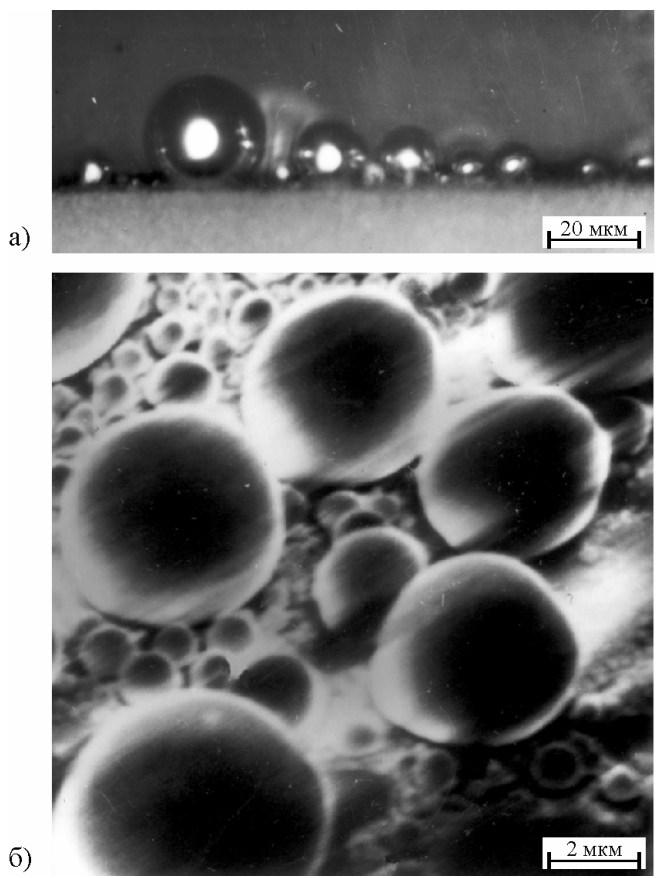

*Рис. 3. Микроснимки, иллюстрирующие измерение краевых углов смачивания методом скола (а – Sn/NaCl) и методом наклонного наблюдения (б – растровая электронная микроскопия, Sn/C) [12]*

Практически удобно измерять величины $2R$, $d$ и $H$ для частиц различного размера, а отношения $H/d$, $d/R$, $H/R$ определять графическим усреднением соответствующих линейных зависимостей, что позволяет минимизировать ошибки измерений параметров частиц. Характерные экспериментальные зависимости $H(R)$, $H(r)$ и $R(r)$ для островковых конденсатов свинца представлены на рис. 4.

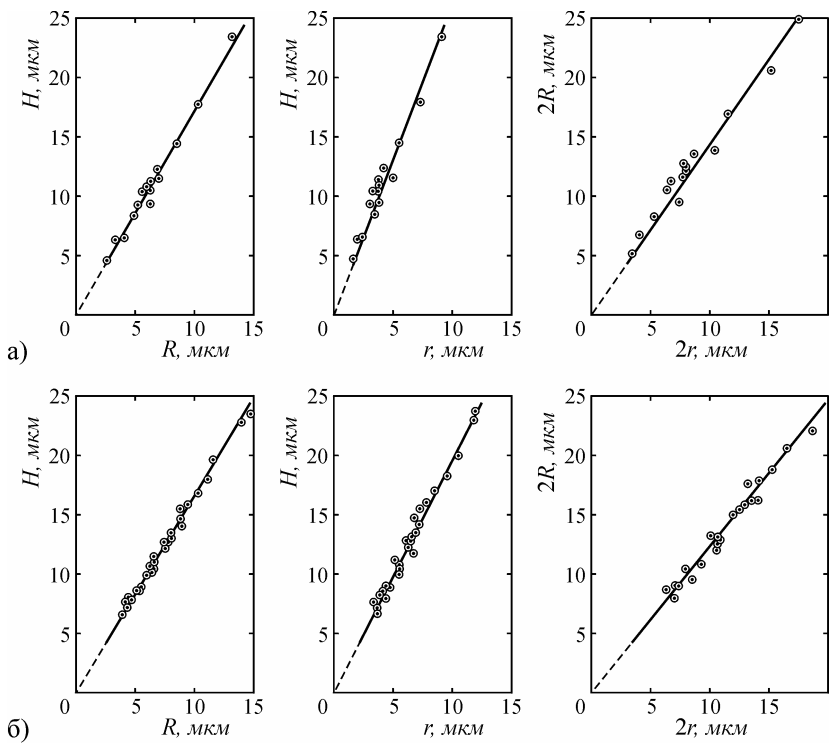

*Рис. 4. Зависимости H(R), H(r) и R(r) для капель свинца на NaCl (а)*
*и углеродных (б) подложках*

Максимальный разброс значений углов, вычисленных по различным соотношениям (2), составляет обычно (2–3)°, что характеризует точность метода. Сферичность формы закристаллизовавшихся капель подтверждается также сравнением измеренной высоты капли со значением, вычисленным через $R$ и $r$ (рис. 5).

Область применимости метода скола для исследований смачивания ограничена размером частиц $R \leq (3–5) \cdot 10^3$ нм, что связано с разрешающей способностью оптического микроскопа. Метод скола накладывает ограничения и на материалы, применяемые в качестве подложек, поскольку подложка должна быть такой, чтобы ее можно было сломать или сколоть, не повредив находящиеся на ней микрочастицы. Анализ показывает, что исследования профиля частиц дают достоверные результаты для систем с ограниченной степенью смачивания ($\theta > 50°$).

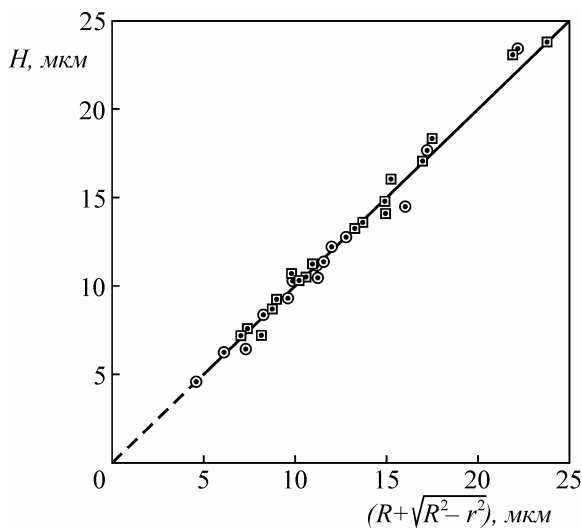

*Рис. 5. Связь между измеренной (H) и вычисленной по R и r высотой*
*(R + (R²–r²)^{1/2}) капель свинца на NaCl (□) и углеродных (○) подложках*

**Метод наклонного наблюдения** [9, 12]. Этот метод определения углов также основан на измерениях параметров профиля капель, но в отличие от метода скола подложка расположена так, что нормаль к ней составляет угол $\alpha$ с оптической осью микроскопа, что дает возможность значительно расширить диапазон как измеряемых углов, так и подложек, поскольку в качестве последних можно использовать любую гладкую поверхность; сечение и вид капель на подложке при наклонном наблюдении представлены на рис. 1б, в. Для нахождения $\theta$ необходимо измерить «кажущуюся высоту» частицы $H_k$ и ее радиус кривизны $R$ ($d$ – при $\theta < 90°$). Угол $\theta$ связан с измеряемыми величинами $R$, $H_k$ и $\alpha$ следующими соотношениями:

$$H_k = R[1 + \sin(\theta - \alpha)] = \frac{d}{2\sin\theta}[1 + \sin(\theta - \alpha)]. \tag{3}$$

Явное выражение $\theta$ через $d$, $H_k$ и $\alpha$ имеет громоздкий вид, поэтому для определения краевых углов при $\theta < 90°$ с помощью этого метода предпочтительно использовать вычислительную технику. Для быстрой оценки $\theta$, как отмечается в работе [9], удобно использовать номограмму, на которой в координатах «$H_k/d$ — $\theta$» нанесено семейство кривых при различных значениях угла наклона подложки (рис. 6).

К достоинствам метода наклонного наблюдения следует отнести наличие значительно большего количества частиц, доступных измерениям, а также применимость его в равной степени для систем с $\theta < 90°$ и $\theta > 90°$. Для повышения точности измерений необходимо, чтобы разность $\alpha - |\theta - 90°|$ составляла не менее $20°$.

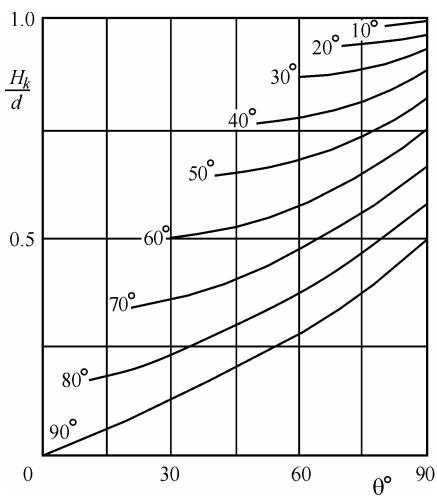

*Рис. 6. Номограмма для расчета краевых углов в методе наклонного наблюдения. Линия 90° соответствует методу скола [9]*

Методы скола и наклонного наблюдения могут быть реализованы также с помощью растровой микроскопии (рис. 3б), применение которой существенно расширяет диапазон разме-ров частиц для исследований смачивания в ультрадисперсных системах.

**Метод зеркального отражения** [12]. Для систем с $\theta < 90°$ разработан метод определения $\theta$, не требующий исследования профиля частиц. Краевой угол определяется через величины $d$ и $R$: диаметр основания $d$ капли на подложке измеряется непосредственно в оптическом микроскопе, а радиус кривизны $R$ находится методом зеркального отражения [10] по изображению удаленного предмета, даваемого сферической поверхностью капли (рис. 7а и рис. 8а).

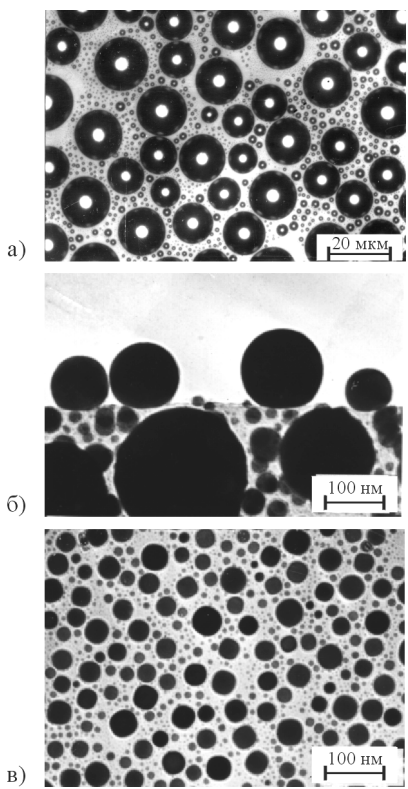

*Рис. 7. Микроснимки к определению краевых углов смачивания методом зеркального отражения (а – островки серы на стеклянной подложке, светлое пятно в центре капель представляет изображение апертурной диафрагмы); методом свертки (б – Pb/Si) и методом фотометрирования (в – Sn/C) [12]*

Из геометрической оптики следует, что радиус кривизны сферической поверхности $R$ может быть определен через расстояние до предмета $S$ и его размер $2l$, которые связаны соотношением

$$R = 2Sr_k/l \quad (\text{при } S \gg R),\tag{4}$$

где $2r_k$ – размер изображения удаленного предмета. Если рассматривать поверхность капли как сферическое зеркало, то из формулы (4) следует, что, измерив $r_k$ при известных $S$ и $l$, можно найти радиус кривизны капли $R$ и по соотношению (2) определить краевой угол. Метод можно использовать для нахождения θ как на прозрачных подложках [10] в проходящем и отраженном свете, так и на непрозрачных подложках в отраженном свете [9]. Частный случай этого метода использован в работе [10] для измерения краевых углов прозрачных жидких капель на прозрачной подложке.

Использование металлографического микроскопа позволяет определять θ для металлических капель на непрозрачных подложках (удаленным предметом может служить апертурная диафрагма микроскопа). Можно показать, что в этом случае выполняется соотношение [9]

$$R = 2r_k F/(AL),\tag{5}$$

где $F$ – фокусное расстояние объектива, $L$ – диаметр апертурной диафрагмы, $A$ – постоянная осветительной системы микроскопа. Из выражений (2) и (5) следует, что краевой угол равен

$$\theta = \arcsin\left(\frac{d}{2r_k}\frac{AL}{2F}\right).\tag{6}$$

Так как $L$ и $F$ известны, то, зная $A$ и измерив $d$ и $r_k$, можно определить θ. Для нахождения постоянной $A$ выполнены измерения величин $R$ и $r_k$ для частиц при θ > 90° (в этом случае $2R = d$) при различных фиксированных значениях диафрагмы $L$ и с разными объективами $F$. Результаты [9] представлены на рис. 8б в координатах «$2r_kF/R$ — $L$», где угловой коэффициент прямой дает значение постоянной $A = 0{,}53$.

Анализ [9, 12] пределов применимости метода показывает, что с его помощью возможно измерять $\theta$ для капель размером $10^3$–$10^5$ нм в интервале углов $40° < \theta < 70°$. Нижний предел обусловлен несферичностью капель при хорошем смачивании, а верхний – увеличением погрешности при приближении $\theta$ к $90°$.

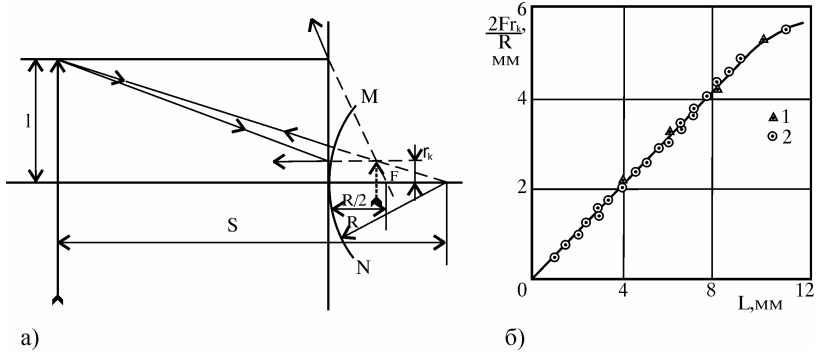

а)                                                                                      б)

*Рис. 8. К методу зеркального отражения [9]: а) ход лучей при определении радиуса кривизны сферической поверхности; б) зависимость величины $2Fr_k/R$ от диаметра апертурной диафрагмы при различных значениях F: 1 – 8,4 мм, 2 – 4,3 мм*

Измерения краевых углов смачивания для частиц размером менее $10^4$ нм проводились с помощью методов электронной микроскопии: метода «свертки» [6] и метода фотометрического анализа [8] электронно-микроскопических снимков капель.

**Метод «свертки»** [6, 8] основан на непосредственном измерении параметров профиля микрочастиц ($R$, $d$, $H$) на электронно-микроскопических снимках, полученных в плоскости, перпендикулярной подложке (рис. 7б). Метод применим для частиц размером $10$–$10^3$ нм. Поскольку для частиц меньших размеров значительно увеличивается относительная погрешность измерений, то для исследований смачивания в сис-темах с $R \leq 10$ нм разработан метод фотометрического анализа электронно-микроскопического изображения частиц.

**Метод фотометрирования** [6, 11, 12, 14]. Угол смачивания θ определяется через параметры $R$ и $r$, которые находятся путем фотометрирования микроснимков частиц, снятых в плане (рис. 7в). На рис. 9а приведена зависимость почернения $S$ от текущей координаты $x$, полученная фотометрированием электронно-микроскопического изображения частицы в диаметральном направлении. Анализ зависимости $S(x)$ позволяет определить параметры $R$ и $r$ следующим образом.

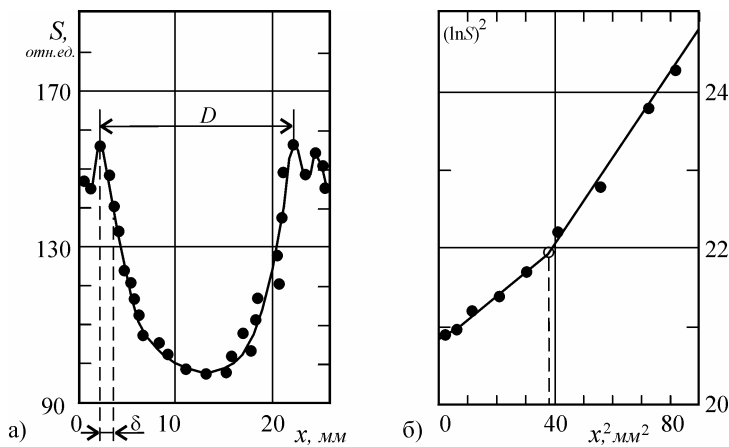

*Рис. 9. Зависимость почернения S от текущей координаты x (а) и зависимость S(x) в координатах «(lnS)²—x²» (б) [12]*

При электронно-микроскопическом исследовании в области изображения края частицы наблюдаются максимумы интенсивности, обусловленные дифракционными эффектами, которые соответствуют максимумам дифракции Френеля. Согласно работе [13], положение френелевских максимумов относительно изображения края объекта зависит от условий фокусировки, так что расстояние от края объекта до $n$-го максимума равно

$$\delta = \left[\Delta f(2n-1)\lambda\right]^{1/2}. \tag{7}$$

Здесь $\Delta f$ – величина дефокусировки объективной линзы, равная расстоянию между фокальной плоскостью объективной линзы и плоскостью, в которой находится объект, $\lambda$ – длина волны электронов. Когда объект находится в фокальной плоскости объектива, величина $\delta = 0$ и краевые контуры не наблюдаются. В электронной микроскопии обычно $n = 1$, поэтому $\delta = [\Delta f\cdot\lambda]^{1/2}$. Величина дефокусировки $\Delta f$ определяется экспериментально с помощью гониометра по углу наклона тест-объекта, представляющего собой коллодиевую пленку с отверстиями диаметром $\sim 10^3$ нм. Угол наклона тест-объекта $\varphi_o$ находится из условий появления френелевского краевого контура при наклоне образца. В этом случае $\Delta f = (L/2)\mathrm{tg}\varphi_0$, где $L$ – линейный размер тест-объекта. Оценки показывают, что при ускоряющем напряжении 100 кВ $\lambda = 3{,}7\cdot10^{-3}$ нм и глубине дефокусировки $\Delta f = 10^2$ нм величина $\delta \approx 0{,}6$ нм.

Таким образом, радиус кривизны частицы определяется по ее электронно-микроскопическому изображению из соотношения

$$R = \tfrac{1}{2}\left(D - 2\delta\right), \tag{8}$$

где $D$ – диаметр первого френелевского контура, который находится фотометрически из зависимости $S(x)$ (рис. 9а). Радиус основания частицы $r$ также определяется из зависимости $S(x)$. На линейном участке денситометрической кривой выполняется соотношение

$$S = kI\tau, \tag{9}$$

где $k$ – коэффициент, учитывающий индивидуальные свойства фотослоя и условия его обработки, $I$ – интенсивность падающего излучения, $\tau$ – время экспозиции. Если в пределах частицы выполняется экспоненциальный закон поглощения энергии, то для интенсивности электронов, прошедших через образец, можно воспользоваться соотношением

$$I = I_0 \exp(-\mu h), \tag{10}$$

где $I_0$ – интенсивность первичного пучка электронов, $\mu$ – коэффициент поглощения. Для частиц, имеющих форму шарового сегмента (рис. 1а), зависимость $h(x)$ существенно различна в двух областях, т. е.

$$h(x) = \begin{cases} a + \left(R^2 - x^2\right)^{1/2} & \text{при } x < r; \\ 2\left(R^2 - x^2\right)^{1/2} & \text{при } r < x < R. \end{cases} \tag{11}$$

Из соотношений (9)–(11) следует, что $\ln S \sim h(x)$, поэтому зависимость $S(x)$ в координатах «$(\ln S)^2 — x^2$» имеет излом при $x = r$ (рис. 9б). Таким образом, на основании данных фотометрирования микроснимков частиц по зависимости $S(x)$ определяются величины $R$ и $r$ для микрочастиц на подложке и, с использованием соотношения (3), вычисляется угол $\theta$.

Описанный метод позволяет находить $\theta$ как для закристаллизовавшихся, так и для жидких частиц с $\theta > 90°$ в диапазоне размеров $3 < R < 70$ нм. Нижний предел размеров определяется чувствительностью микрофотометра, а верхний предел связан с предельной толщиной объекта, которая еще прозрачна для электронов.

Метод можно распространить и на случай $\theta < 90°$. Действительно [12], для частицы, имеющей форму шарового сегмента с радиусом кривизны $R$ и высотой $H$ (рис. 1а), зависимость $h(x)$ имеет вид

$$h(x) = H - R + \left(R^2 - x^2\right)^{1/2}. \tag{12}$$

Если диаметр основания частицы меньше экстинционной длины (условие отсутствия толщинных контуров на электронно-микроскопическом изображении частиц), то контраст изображения определяется только толщиной поглощающего слоя $h$. С использованием формул (9), (10) и (12) получается

$$\ln S(x) = \ln S_0 - \mu\left[H - R + \left(R^2 - x^2\right)^{\frac{1}{2}}\right]. \qquad (13)$$

В общем случае, когда коэффициент поглощения $\mu$ неизвестен, для нахождения $\theta$ требуется решить систему из $n$ условных уравнений ($n$ – число экспериментальных точек) с неизвестными $\mu$ , $H$ и $R$. Решение можно найти методом наименьших квадратов, определив ранее приближенные значения $\mu_0$ , $H_0$ и $R_0$ и разложив выражение (13) в ряд по степеням $\Delta\mu = \mu - \mu_0$, $\Delta H = H - H_0$, $\Delta R = R - R_0$. Пренебрегая членами выше 1-го порядка, получается система уравнений для поправок $\Delta\mu, \Delta H, \Delta R$:

$$Y_i = \Delta H + \Delta R \cdot X_i + M \cdot Z_i, i = 1,2,...n , \qquad (14)$$

где $Y_i = y_i/\mu_0$ , $X_i = R_0\left/\left(R_0^2 - x^2\right)^{\frac{1}{2}}\right. - 1$ ,

$Z_i = H_0 - R_0 + \left(R_0^2 - x^2\right)^{\frac{1}{2}}$ , $M = \Delta\mu/\mu_0 + 1$ .

Решение системы уравнений (14) для нахождения неизвестных $M, \Delta H$ и $\Delta R$ может быть получено методом наименьших квадратов. В приведенном способе обработки экспериментальных данных для определения угла $\theta$ при $\theta < 90°$ предполагалось, что коэффициент поглощения электронов $\mu$ неизвестен. Метод фотометрирования позволяет независимым способом определить коэффициент $\mu$. Действительно, из формулы (13) следует, что, зная зависимость $h(x)$ и представив данные по фотометрированию в координатах «$\ln S - h$», можно определить $\mu$. Для этого удобно использовать результаты фотометрирования микроснимков профилей частиц с $\theta > 90°$, поскольку при этом исключается поглощение электронов в подложке. Для частиц с $\theta > 90°$ толщина поглощающего слоя равна $h(x) = 2\left(R^2 - x^2\right)^{\frac{1}{2}}$. Определив $\mu$ из наклона зависимостей $S(h)$ в координатах «$\ln S - h$» (рис. 10) для исследуемого металла, можно существенно упростить анализ зависимости для нахождения угла смачивания в системах с θ < 90° [12].

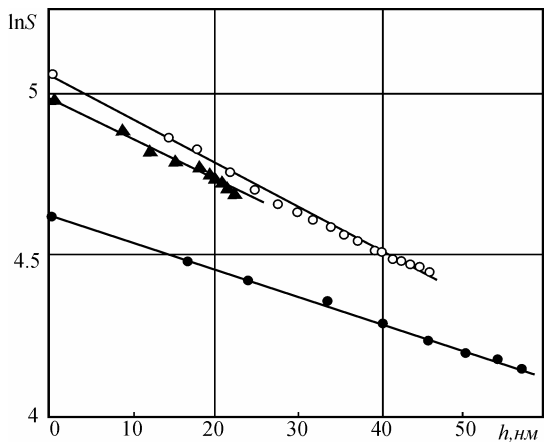

*Рис. 10. К определению коэффициента поглощения электронов. Зависимости S(h), полученные фотометрированием профилей микрочастиц олова (●), свинца (○) и висмута (▲) (ускоряющее напряжение 100 кВ)*

**Оценка θ по гистограммам** [12]. Этот метод дает возможность определить усредненное значение краевого угла смачивания для ансамбля частиц, составляющих островковую пленку. Пусть островковая пленка состоит из частиц, представляющих собой подобные сферические сегменты с радиусом кривизны в интервале *0 < R < R_max*. Значение $R_{max}$ определяется количеством сконденсированного вещества. Масса единицы площади такой пленки равна

$$m = \rho \int\limits_{0}^{R_{\max}} V(R,\theta) \cdot n(R) dR,\qquad(15)$$

где ρ – плотность, $V(R,\theta) = \frac{4}{3}\pi R^3 \cdot \Phi(\theta)$ – объем частицы, *n(R)* – функция распределения частиц по размерам.

Предполагая, что θ не зависит от *R* (это условие выполняется для частиц с *R > 30* нм), и определив величины *m* и *n(R)*, можно найти функцию Φ(θ) для островковой пленки, т. е.

$$\Phi(\theta) = \frac{3m}{4\pi\rho} \left( \int\limits_0^{R_{\max}} n(R)R^3 dR \right)^{-1}, \tag{16}$$

где $\Phi(\theta) = \frac{1}{4}(2+\cos\theta)(1-\cos\theta)^2$. В случае $\theta < 90°$ измеряется радиус основания сегмента $r = R\sin\theta$ и соотношение (16) преобразуется к виду

$$\frac{\Phi(\theta)}{\sin^3\theta} = \frac{3m}{4\pi\rho} \left( \int\limits_0^{r_{\max}} n(r)r^3 dr \right)^{-1}. \tag{17}$$

Следовательно, зная массу сконденсированного вещества и построив гистограмму по снимку островковой пленки, полученной конденсацией по механизму пар→жидкость, можно определить усредненное значение $\theta$ для ансамбля частиц.

**Другие методы** [12]. В ряде случаев, особенно при использовании оптической микроскопии, возможно применение других методик, отличающихся способом измерения характерных геометрических параметров покоящейся капли. Так, для «больших» частиц ($H > 10^4$ нм) угол $\theta$ может быть определен через параметры $R$ и $H$ (или $d$ и $H$ в случае $\theta < 90°$), которые находятся следующим образом. Величина $R$ (или $d$) измеряется непосредственно при исследовании плана подложки в оптическом микроскопе, а высота капли определяется по величине перемещения объектива при перефокусировке с поверхности подложки (основания капли) на вершину капли.

При исследовании смачивания прозрачных подложек величина $\theta$ ($\theta > 90°$) может быть найдена из сравнения микроснимков обратной стороны участка островковой пленки, полученных в проходящем и отраженном свете. В первом случае измеряются диаметры частиц ($2R$), а во втором – соответствующие диаметры оснований ($d$).

Таким образом, разработанный комплекс экспериментальных методов [6–12] позволяет исследовать смачивание поверхностей твердых тел малыми металлическими частицами при изменении размеров последних в пределах $3–10^5$ нм.

*Глава 2*

# Размерный эффект при смачивании

Смачивание в системе «жидкость – твердое тело» характеризуется величиной равновесного краевого угла $\theta$, который связан с поверхностными энергиями контактирующих фаз уравнением Юнга

$$\cos\theta = \left(\sigma_u - \sigma_{ul}\right)/\sigma_l \, , \qquad (18)$$

где индексы $u$ и $l$ относятся к твердой (подложке) и жидкой (частице) фазе соответственно. Можно ожидать, что размерная зависимость поверхностной энергии будет вызывать отличия закономерностей смачивания в нанодисперсных системах от известных для макроскопических объектов. При этом, в частности, возможно изменение краевого угла с увеличением степени дисперсности как жидкой, так и твердой фаз. Для описания таких явлений необходимо решить задачу о равновесной форме микрокапли и ее краевом угле, т. е. получить аналог уравнения Юнга (18) с учетом зависимости $\sigma(R)$.

## 2.1. Угол смачивания с учетом размерной зависимости поверхностной энергии

Рассмотрим, следуя [15], малую каплю жидкости, находящуюся на плоской твердой поверхности. Полная свободная энергия системы $F$ состоит из гидростатической энергии $pV$ (здесь давление $p$ можно рассматривать как неопределенный множитель Лагранжа, учитывающий постоянство объема капли $V$) и энергии поверхностных сил

$$F = -pV + \int\limits_{S_l} \sigma_l dS + \int\limits_{S_{ul}} (\sigma_{ul} - \sigma_u) dS, \qquad (19)$$

где $S$ – площадь соответствующей поверхности раздела.

Поверхностная энергия $\sigma_l$, в соответствии с существующими представлениями [15, 16], предполагается зависящей от средней кривизны поверхности $C$ в данной точке:

$$\sigma_l = \sigma_l^\infty (1 - \alpha C). \qquad (20)$$

Для сферической поверхности $(C = 1/R)$ зависимость (20) совпадает с выражением $\sigma_l = \sigma_l^\infty (1 - \alpha/R)$ [16], известным в литературе как формула Толмена.

При нахождении условий равновесия необходимо также учитывать размерную зависимость межфазной энергии границы капля – подложка $\sigma_{ul}$ [8]. В работах [8, 11, 17] предполагалось, что эта зависимость описывается уравнением $\sigma_{ul} = \sigma_{ul}^\infty (1 - \beta/R)$, однако, более естественно, как это сделано в [15, 18], принять зависимость межфазной энергии не от радиуса кривизны поверхности жидкости $R$, а от радиуса периметра смачивания $r$, т. е. воспользоваться для $\sigma_{ul}$ зависимостью вида

$$\sigma_{ul} = \sigma_{ul}^\infty (1 - \beta/r). \qquad (21)$$

Выражения (20) и (21) применимы при $1/C \gg \alpha$ и $r \gg \beta$. Отметим также, что при нахождении равновесных характеристик капли не делается никаких допущений о знаке и величине параметров $\alpha$ и $\beta$.

Впервые размерный эффект при смачивании теоретически рассматривался в работах [19, 20], авторы которых ограничились учетом только зависимости $\sigma_l(R)$ в виде (20) и получили для краевого угла (случай $\theta < 90°$) малых частиц выражение

$$\cos\theta = \frac{\cos\theta_\infty}{1-2\alpha/R} \qquad \left(\cos\theta_\infty = \left(\sigma_u - \sigma_{ul}^\infty\right)/\sigma_l^\infty\right). \qquad (22)$$

Здесь $\theta_\infty$ – значение угла при $R \to \infty$. Так как $\alpha > 0$, то из выражения (22) следует, что для контактных систем с $\theta < 90°$ размерный эффект должен проявляться в уменьшении $\theta$ с размером частиц (это качественно согласуется с экспериментами [21]), а при $\theta > 90°$ краевой угол должен увеличиваться с уменьшением $\theta$. Этот вывод противоречит экспериментальным результатам работ [8, 14], а также представлениям о кинетике зарождения и роста островковых вакуумных конденсатов на подложках. Например, известно, что материал подложки вносит существенный вклад в процессы зарождения и роста частиц, влияя на температурную стабильность переохлажденной жидкой фазы в островковых пленках [77, 78]. Однако из выражения (22) следует, что для очень малых частиц (стадия возникновения зародышей) $\theta \to 180°$ и подложка не должна влиять на образование островков.

Неоднозначность такого подхода показана в [8], где с учетом зависимости межфазной энергии от радиуса кривизны капли получено выражение для краевого угла смачивания в виде

$$\cos\theta = \frac{\sigma_u - \sigma_{ul}^\infty\left(1-2\beta/R\right)}{\sigma_l^\infty\left(1-2\alpha/R\right)}. \qquad (23)$$

Однако, как показал анализ результатов этих работ [15, 18], полученные в [8, 19] выражения являются неточными вследствие использования при расчетах соотношений вида $S\,\delta R = -R\,\delta S$ ($\delta R,\ \delta S$ – вариации радиуса и площади поверхности капли), в которых не учтено изменение формы капли при варьировании.

В связи с осевой симметрией задачи для ее решения [15] можно воспользоваться полярными координатами с началом в центре окружности периметра смачивания и вертикальной осью $z$, перпендикулярной плоскости подложки. Профиль сво-бодной поверхности капли задается функцией $z(\rho)$. Без нарушения общности можно считать функцию $z(\rho)$ однозначной, т. е. рассмотреть случай $\theta < 90°$ (можно показать, что полученные результаты будут справедливы во всем интервале углов $\theta$, если выполнить более громоздкие преобразования, выбрав в качестве независимой переменной $z$ и задав поверхность капли однозначной при любых $\theta$ функцией $\rho(z)$).

Равновесная форма капли находится минимизацией функционала (19), который с учетом соотношений для объема капли и площадей ограничивающих ее поверхностей записывается следующим образом:

$$F = 2\pi\int\limits_0^r\left[-pz + \sigma_l(C)\left(1 + z'^2\right)^{1/2} + \sigma_{ul}(r) - \sigma_u\right]\rho d\rho,$$

$$C = -\frac{1}{2}\left\{\frac{z''}{\left(1 + z'^2\right)^{3/2}} + \frac{z'}{\rho\left(1 + z'^2\right)^{1/2}}\right\}. \tag{24}$$

Слагаемое $(\sigma_{ul}(r) - \sigma_u)\rho$ в подынтегральном выражении не содержит $z(\rho)$ и ее производных, т. е. зависимость $\sigma_{ul}(r)$ определяет только граничные условия и не влияет на форму капли.

Варьирование функционала (24) по $\delta z$ приводит к уравнению Эйлера, которое после почленного интегрирования принимает вид

$$-\frac{p\rho^2}{2} = \rho\left(1 + z'^2\right)^{1/2}\frac{d\sigma_l}{dC}\frac{\partial C}{\partial z'} + \frac{\sigma_l\rho z'}{\left(1 + z'^2\right)^{1/2}} - \frac{d}{d\rho}\left[\rho\left(1 + z'^2\right)^{1/2}\frac{d\sigma_l}{dC}\frac{\partial C}{\partial z''}\right]. \tag{25}$$

Постоянная интегрирования в (25) оказывается равной нулю из условия равенства нулю одного из неинтегральных слагаемых $\delta F$ в точке $\rho = 0$. Сложность уравнения (25) делает маловероятным его общее решение, что вынуждает прибегнуть к конкретизации зависимости $\sigma_l(C)$ в виде (20). Подстановка в (25) соотношения (20) и выражений для производных $d\sigma_l/dC$, $\partial C/\partial z'$ и $\partial C/\partial z''$ приводит к нелинейному дифференциальному уравнению первого порядка

$$\left[ z'\left(1+z'^2\right)^{-\frac{1}{2}}\right]^2 + \frac{2\rho}{\alpha}\left[ z'\left(1+z'^2\right)^{-\frac{1}{2}}\right] + \frac{p\rho^2}{\alpha\sigma_l} = 0\,, \qquad (26)$$

решение которого путем разделения переменных дает равновесную форму поверхности капли в виде сферы, усеченной плоскостью $z = 0$:

$$\left(z - z_0\right)^2 + \rho^2 = R^2\,. \qquad (27)$$

Радиус сферы удовлетворяет соотношению

$$p = \frac{2\sigma_l^\infty}{R}\left(1 - \frac{\alpha}{2R}\right), \qquad (28)$$

из которого видно, что неопределенный множитель Лагранжа $p$ есть не что иное, как лапласово давление с поправкой на зависимость $\sigma(R)$. Постоянная интегрирования $z_0 = \pm\sqrt{R^2 - r^2}$ имеет смысл $z$-координаты центра сферы (27) и определяется из условия $z(r) = 0$.

Угол смачивания $\theta$ можно найти либо из граничных условий, либо, так как функция $z(\rho)$ определена, из условия минимума свободной энергии капли при постоянном объеме.

Прежде чем перейти к вычислениям, возвратимся к уравнению (25), полученному без каких-либо предположений о виде зависимости $\sigma_l(C)$. Легко убедиться, что функция $z(\rho)$ в виде (27) является его частным решением, при этом для $p$ следует известное в литературе выражение $p = \dfrac{2\sigma_l}{R} + \dfrac{d\sigma_l}{dR}$ для лапласова давления с учетом влияния кривизны на поверхностную энергию [22]. Полученное соотношение (28) является его частным случаем для зависимости $\sigma_l(R)$ в виде (20). Отметим также, что сферическая форма капли при монотонной зависи-мости $\sigma_l(R)$ с точки зрения физики рассматриваемого явления выглядит наиболее естественной.

Приведенные рассуждения [15] не следует рассматривать как строгое доказательство, однако они дают право обоснованно предположить сферичность свободной поверхности капли и вернуться к решению задачи в общем виде, т. е. с произвольными $\sigma_l(R)$ и $\sigma_{ul}(R)$.

Для нахождения краевого угла как функции $R$ необходимо выразить поверхностную энергию капли через $R$ и $\theta$:

$$F_s = \pi R^2 \left\{ 2\sigma_l(R)(1 - \cos\theta) + \left[ \sigma_{ul}(r) - \sigma_u \right] \sin^2\theta \right\} \qquad (29)$$

Приравнивая нулю производную $dF_s/dR$, можно получить уравнение, которое при учете постоянства объема приводит к условию равновесия микрокапли на подложке

$$\cos\theta = \frac{\sigma_u - \sigma_{ul} - R\dfrac{d\sigma_l}{dR} - \dfrac{r}{2}\dfrac{d\sigma_{ul}}{dr}}{\sigma_l + R\dfrac{d\sigma_l}{dR}}. \qquad (30)$$

Уравнение (30), как и следовало ожидать, отличается от уравнения Юнга (18) наличием слагаемых, содержащих производные поверхностной энергии по размеру.

Используя выражения (20) и (21) для $\sigma_l(R)$ и $\sigma_{ul}(R)$, можно записать соотношение для краевого угла микрочастицы через параметры $\alpha$ и $\beta$, определяющие размерную зависимость соответствующих поверхностных энергий, следующим образом [15]

$$\cos\theta = \cos\theta_\infty - \frac{\alpha}{R} + \frac{\beta}{2R}\frac{\sigma_{ul}^\infty}{\sigma_l^\infty}\frac{1}{\sin\theta}. \qquad (31)$$

Естественно, что в предельном случае при $\sigma \to \sigma^\infty$ ($\alpha/R \to 0$, $\beta/R \to 0$) все полученные выражения переходят в известные соотношения теории капиллярности.

## 2.2. Размерные зависимости угла смачивания и энергии поверхности раздела твердое тело – жидкость

Первые экспериментальные данные об изменении краевого угла смачивания с уменьшением размеров жидких капель были получены в работе [21], в которой электронно-микроскопически исследовалось смачивание малыми каплями диффузионного вакуумного масла и гидрооксихлорида титана тонких нитей асбеста и окиси ванадия (эти контактные системы соответствуют случаю $\theta < 90°$). Было установлено, что угол $\theta$ зависит как от размера капли $R$, так и от радиуса нити $r_u$, причем улучшение смачивания происходит при уменьшении $R$ или увеличении $r_u$. Это явление наблюдалось для исследованных контактных систем при размерах капель менее 1000 нм и нитей диаметром менее 200 нм. Полученные в работе [21] результаты указывают на размерную зависимость поверхностных энергий контактирующих фаз, однако, вследствие сложности рассматриваемой системы (наличие двух переменных параметров $R$ и $r_u$), имеющихся данных недостаточно для количественного анализа.

Теоретически вопрос о смачивании малой каплей тонкой нити рассматривался в работах [5, 23], где из условий минимума свободной энергии системы с учетом размерных зависимостей поверхностной энергии капли $\sigma_l(R)$ и адгезионного натяжения $\sigma_u - \sigma_{ul} = f(r_u)$ было получено уравнение, связывающее равновесное значение краевого угла с радиусами капли и нити

$$\frac{\cos\theta}{\cos\theta_\infty} = \frac{1 - \alpha/2r_u}{1 - 2\alpha/R}. \tag{32}$$

Здесь $\alpha$ – параметр, определяющий размерную зависимость поверхностной энергии жидкой фазы, а $\theta_\infty$ – значение угла смачивания при $R \to \infty$ и $r_u \to \infty$.

Анализ, выполненный авторами работ [5, 23], показывает, что при $\alpha > 0$ смачивание должно улучшаться как с уменьшением размера капли, так и при увеличении радиуса нити, то есть соотношение (32) качественно верно описывает обнаруженное экспериментально в работе [21] изменение краевого угла. Совпадение экспериментальных данных [21] с теоретическими следствиями, полученными в предположении $\alpha > 0$ [5, 23], свидетельствует о понижении поверхностной энергии жидкой капли с уменьшением ее размера.

Экспериментальные работы, посвященные изучению зависимости $\theta(R)$ для микрокапель на плоской подложке, появились сравнительно недавно, когда были разработаны надежные электронно-микроскопические методы измерения краевых углов.

Впервые размерный эффект при смачивании малыми металлическими каплями поверхности плоской твердой подложки экспериментально был обнаружен для островковых вакуумных конденсатов олова и индия на аморфной углеродной подложке [8, 14]. Островковые пленки получались путем испарения и конденсации олова и индия по механизму пар→жидкость на аморфных углеродных подложках в вакууме $1\cdot10^{-7}$ мм рт. ст., создаваемом при помощи сорбционных угольных насосов и геттероионного насоса типа «орбитрон». Сочетание методов оптической и электронной микроскопии [12] позволило определить краевые углы смачивания частиц радиусом $(1–10^4)$ нм.

Измерения краевых углов смачивания для частиц микронных размеров выполнялись следующим образом. На скол (001) монокристалла NaCl в вакууме напылялся сплошной слой углерода, а затем без нарушения вакуума по механизму пар→жидкость конденсировался исследуемый металл (температура подложки в случае олова 520 К); при этом на подложке образуются жидкие частицы, наиболее вероятный размер которых определяется ко-личеством сконденсированного металла. Полученные образцы охлаждались в вакууме до комнатной температуры, и затем подложки с закристаллизовавшимися частицами исследовались в оптическом микроскопе. Применение в качестве подложкодержателей монокристаллов NaCl позволяло делать поперечные сколы и измерять параметры профиля частиц. Микроскопические измерения параметров профиля частиц размером $(2–25) \cdot 10^3$ нм указывают на сферичность их формы.

Согласно проведенным измерениям [8], краевой угол для системы Sn/C(пленка) в указанном интервале размеров частиц постоянен и равен $151° \pm 2°$, что согласуется с известными данными для микрокапель олова на аморфном углероде [1].

Образцы для электронно-микроскопических исследований препарировались аналогичным способом. Для измерения краевого угла применялись метод «свертки» и метод фотометрического анализа снимков [12]. Эти измерения показали, что частицы размером 10–100 нм также имеют форму шарового сегмента и для частиц радиусом $R > 30$ нм величина угла $\theta$ близка к значению для частиц микронных размеров. Так как точность измерений $\theta$ методом свертки понижается с уменьшением размеров частиц, то дальнейшие измерения были выполнены с помощью специально разработанного для этих целей метода, основанного на фотометрическом анализе электронно-микроскопических снимков на просвет.

Результаты измерений $\theta$ для олова на углеродной подложке приведены на рис. 11а, из которого видно, что для больших частиц ($R > 30$ нм) значения, полученные методом свертки и методом фотометрирования, хорошо согласуются между собой и близки к соответствующим величинам для частиц микронных размеров. При уменьшении размеров частиц ($R < 30$ нм) наблюдается уменьшение краевого угла.

Позже в работах [11, 17, 24, 26, 27] были выполнены исследования смачивания в системах «островковая пленка металла (Bi, Pb, Au) – аморфная пленка углерода» и «Pb – аморфная пленка кремния» в зависимости от размера частиц. В указанных системах, как и в случае олова и индия на углеродной подложке, практически отсутствует растворимость в твердом и жидком состояниях, т. е. подложка по отношению к исследуемым металлам является изотропной и химически нейтральной. В различных экспериментах вакуум при препарировании пленок находился в пределах $10^{-7}$–$10^{-9}$ мм рт. ст.

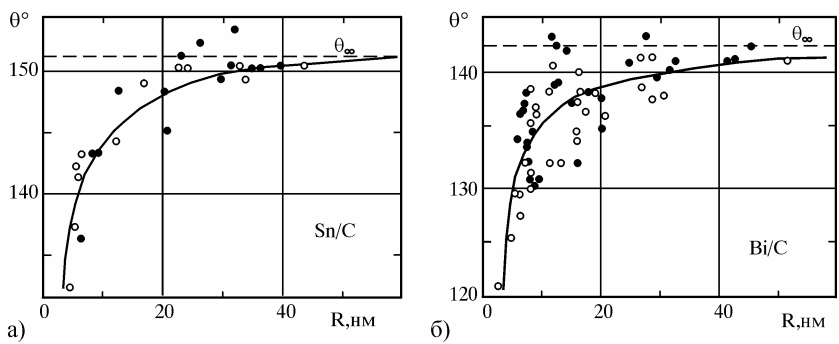

*Рис. 11. Зависимости краевого угла смачивания от радиуса частиц олова (а) [8] и висмута (б) [11] на аморфной углеродной подложке, полученные методом свертки (○) и методом фотометрирования (●)*

Для всех исследованных систем получено, что при размерах частиц $R > 100$ нм и толщинах аморфных пленок углерода и кремния $t > 20$ нм, краевые углы смачивания жидких капель в островковых пленках хорошо согласуются с данными для соответствующих контактных систем в массивном состоянии [1]. При размерах частиц $R < 30$ нм величина $\theta$ уменьшается, так что $\Delta\theta = \theta_\infty - \theta \approx (20°–25°)$ при $R = (4–5)$ нм (рис. 11, 12).

Следует отметить, что смачивание микрочастицами золота и серебра углеродных подложек исследовалось электронно-микроскопически методом свертки в работе [29]. Образцы для исследований препарировались конденсацией в вакууме. Для уменьшения плотности островков на подложке применялась электрокинетическая коалесценция в среде HCl и $H_2SO_4$. Было установлено, что для частиц размером 5–100 нм величины $\theta$ не отличаются от соответствующих значений для массивных образцов. Анализ использованной в работе [29] методики получения и последующей обработки образцов показывает [17], что этот результат, вероятно, обусловлен тем, что измеренные углы смачивания не соответствуют своим равновесным значениям. Это связано с тем, что образование частиц происходило путем твердофазной коалесценции, о незавершенности которой свидетельствуют перешейки между частицами, наблюдаемые на приведенных в работе [29] электронно-микроскопических снимках.

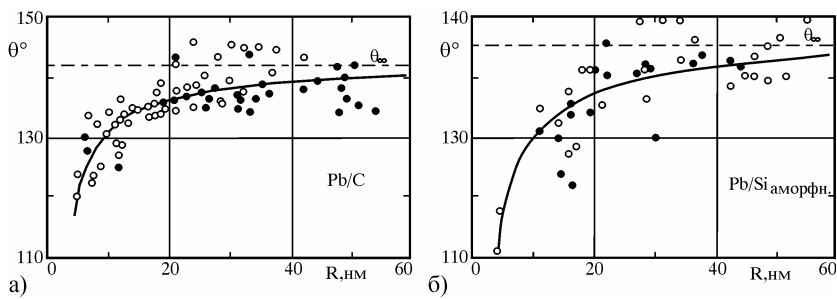

*Рис. 12. Зависимость краевого угла смачивания от радиуса частиц для островковых конденсатов свинца на аморфных углеродной (а) [11] и кремниевой (б) [17] подложках (○– данные по методу свертки, ● – по фотометрированию электронно-микроскопических снимков)*

Известно, что растворенные или адсорбированные на поверхности примеси могут оказывать существенное воздействие на характер физико-химических процессов на границе раздела фаз [2, 3, 30]. Для высокодисперсных систем примесные эффекты должны проявляться в большей мере, поскольку в них с уменьшением характерного размера наблюдается увеличение растворимости по сравнению с массивными образцами. Усло-вия препарирования являются технологическим параметром, определяющим в значительной мере содержание примесей в конденсированных пленках. По этой проблеме, как уже неоднократно отмечалось, накоплен обширный фактический материал. Однако смачивание в высокодисперсных системах в этом плане практически не исследовалось, и поэтому представлялось важным [27] определить влияние условий препарирования образцов на краевые углы микрочастиц с целью выяснить, когда определяющим будет размерный эффект. Для решения этих задач в работе [27] были выполнены исследования смачивания углеродных пленок микрочастицами золота, конденсированными в существенно различных вакуумных условиях ($10^{-3}$ мм рт. ст. и $3 \cdot 10^{-9}$ мм рт. ст.), создаваемых при помощи безмасляных средств откачки.

В вакууме образование на поверхности подложки монослоя остаточных газов при условии равенства коэффициента конденсации единице при давлении $10^{-3}$ мм рт. ст. (индекс $lo$) и $10^{-9}$ мм рт. ст. (индекс $hi$) происходит в течение $\tau_{lo} = 10^{-3}$ с и $\tau_{hi} = 10^3$ с соответственно. Время препарирования островковых пленок золота [17, 27] составляло $\tau \leq 5$ с, т. е. $\tau_{lo} \ll \tau \ll \tau_{hi}$. Как показывают оценки [27], при конденсации в вакууме $10^{-3}$ мм рт. ст. атомы золота и атомы примесей из остаточных газов поступают на подложку в отношении 1:100, а в вакууме $10^{-9}$ мм рт. ст. это отношение составляет $10^4$:1. Следовательно, островки золота, полученные в указанных условиях, должны содержать существенно различное количество растворенных газовых примесей (конденсация осуществлялась через жидкую фазу), при этом в вакууме $10^{-9}$ мм рт. ст. их общее количество в принципе не может превышать 0,01 ат.%. Поскольку островковые пленки золота [17, 27] препарировались при температуре углеродной подложки ~1350 К, при которой коэффициент конденсации для большинства остаточных газов очень мал, то реальное содержание примесей будет существенно меньше, чем 0,01 ат.%.

Результаты измерений [17, 27] краевых углов смачивания в системе Au/C приведены на рис. 13. Как в вакууме $3 \cdot 10^{-9}$ мм рт. ст. ($\theta^{hi}$), так и в вакууме $10^{-3}$ мм рт. ст. ($\theta^{lo}$) наблюдается размерный эффект смачивания, проявляющийся в понижении краевого угла смачивания с уменьшением радиуса капель при $R < 30$ нм, а при $R > 40$ нм зависимости $\theta^{hi}(R)$ и $\theta^{lo}(R)$ выходят на постоянное значение $\theta_{\infty}$.

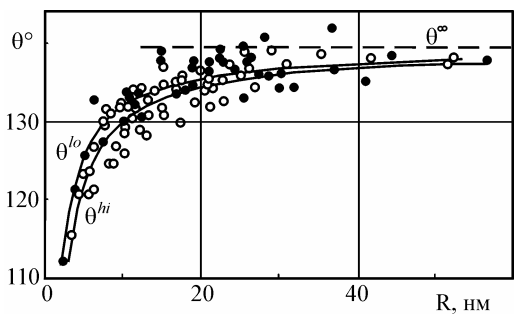

*Рис. 13. Зависимость $\theta(R)$ для микрокапель золота, конденсированных в вакууме $3 \cdot 10^{-9}$ (○) и $10^{-3}$ (●) мм рт. ст. на углеродной подложке [12, 27]*

Из рис. 13 видно, что характер обеих зависимостей в основном такой же, как и для рассмотренных выше контактных систем. Однако, полученные [27] данные указывают, что при фиксированном размере частиц значения краевых углов при конденсации в вакууме $10^{-3}$ мм рт. ст. несколько больше, чем в вакууме $3 \cdot 10^{-9}$ мм рт. ст.

Для анализа этих результатов в работе [27] было использовано условие равновесия малой капли на подложке (31), записанное в первом приближении в следующем виде:

$$\cos \theta = \cos \theta_{\infty} + \gamma / R. \tag{33}$$

Параметры $\theta_{\infty}$ и $\gamma$ были найдены путем обработки экспериментальных зависимостей $\theta^{hi}(R)$ и $\theta^{lo}(R)$ методом наимень-ших квадратов и составили при конденсации в вакууме $10^{-3}$ мм рт. ст. $\theta_\infty = 139{,}4°$ и $\gamma = 0{,}88$ нм, а в вакууме $3 \cdot 10^{-9}$ мм рт. ст. $\theta_\infty = 139{,}7°$ и $\gamma = 1{,}14$ нм. На рис. 13 представлены зависимости $\theta^{hi}(R)$ и $\theta^{lo}(R)$, аппроксимированные уравнением (33) с использованием соответствующих величин $\theta_\infty$ и $\gamma$. Из сравнения этих зависимостей следует, что $\theta^{hi} > \theta^{lo}$ и разность $\Delta\theta = \theta^{lo} - \theta^{hi}$ также изменяется с размером частиц (рис. 14а).

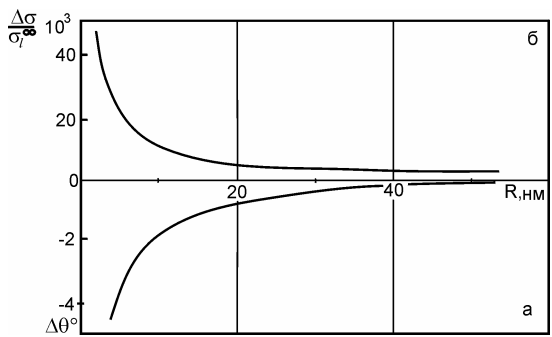

*Рис. 14. Зависимость разности углов смачивания $\Delta\theta = \theta^{lo} - \theta^{hi}$ (а) и поверхностных энергий жидкой фазы $\Delta\sigma = \sigma_l^{lo} - \sigma_l^{hi}$ (б) от размеров частиц для островковой пленки золота на углеродной подложке [27]*

Зависимость $\Delta\theta(R)$ объясняется [27] следующим образом. Если учесть, что островки золота конденсировались при высоких температурах подложки ($\sim 1350$ K), то основное влияние должны оказывать растворенные газовые примеси (так как адсорбция газов при таких температурах пренебрежимо мала). Следовательно, воздействие остаточных газов на угол смачивания должно проявляться через величину $\sigma_l$. Это подтверждается полученными данными, так как величины $\theta_\infty^{lo}$ и $\theta_\infty^{hi}$ практически совпадают, в то время как параметры $\gamma^{lo}$ и $\gamma^{hi}$ оказываются различными. Таким образом, величина давления остаточных газов при препарировании островковых пленок золота незначительно сказывается на значениях поверхностных энергий, соответствующих массивному состоянию (золото, как известно, инертный к остаточным газовым примесям металл), но влияет на размерную зависимость поверхностной энергии.

С использованием экспериментальных зависимостей $\theta(R)$ и $\Delta\theta(R)$ в [27] было определено изменение поверхностной энергии $\sigma_l$, обусловленное растворенными газовыми примесями. Полученная [27] зависимость $\dfrac{\Delta\sigma}{\sigma_l^\infty}(R)$ представлена на рис. 14б, из которого следует, что с уменьшением размеров частиц $\Delta\sigma = \sigma_l^{hi} - \sigma_l^{lo}$ возрастает. Поскольку поверхностная энергия металлов понижается с увеличением концентрации растворенных газовых примесей, полученные результаты [27] указывают на увеличение растворимости примесей из остаточных газов с уменьшением размера частиц золота.

Как показывают результаты исследований смачивания в системе Au/C, свойства островковых пленок, даже такого слабо чувствительного к влиянию остаточных газов металла, как золото, зависят от условий их получения, даже в тех случаях, когда изменение этих условий не сказывается на соответствующих характеристиках массивных объектов. Тем не менее даже при конденсации в низком вакууме сохраняется характер зависимости $\theta(R)$. Поэтому можно заключить, что приведенные на рис. 11–13 данные по смачиванию в различных системах, полученные в вакууме $10^{-7}$–$10^{-9}$ мм рт. ст. при безмасляной откачке, свидетельствуют о размерном эффекте смачивания, а не являются следствием влияния примесей.

Результаты исследований смачивания, полученные в [8, 11, 14, 17] и приведенные на рис. 11–13, представляют самостоятельный интерес, в частности для некоторых практических приложений, но наряду с этим позволяют получить новую физическую информацию о свойствах микрочастиц.

В соответствии с результатами разделов 2.1 и 2.2 по экспериментальным зависимостям $\theta(R)$ и $\sigma_l(R)$ можно определить размерную зависимость межфазной энергии границы микрочастица – подложка $\sigma_{ul}$.

Для анализа результатов по размерному эффекту смачивания в островковых пленках использовано [18] выражение (31), из которого по зависимости $\theta(R)$, при известных значениях параметра $\alpha$ и величине поверхностной энергии $\sigma_u$, определены величина межфазной энергии границы – жидкая микрочастица-подложка и ее размерная зависимость в соответствии с (21), т. е. параметр $\beta$. Параметр $\alpha$ может быть найден из данных по кинетике испарения малых частиц [52]. Его можно оценить также, воспользовавшись соотношением $\alpha \approx 0{,}91(v_a/N_a)$ ($v_a$ – атомный объем, $N_a$ – число Авогадро) [22]. Расчет $\alpha$ по этому соотношению дает значение $\alpha(Pb) = 0{,}58$ и $\alpha(Au) = 0{,}48$ нм, т. е. величины, близкие к найденным экспериментально в работе [52]. Это позволяет в первом приближении использовать указанное соотношение для оценок параметра $\alpha$ металлов, для которого нет экспериментальных данных по зависимости $\sigma_l(R)$. Величина поверхностной энергии углеродной пленки была определена в работе [55] из данных по смачиванию микрокаплями индия, олова и свинца свободных пленок различной толщины и составляет $\sigma_u = 120 \pm 30$ мДж/м$^2$.

Используя эти данные, в работе [18] были определены величины $\sigma_{ul}$ и параметр $\beta$, которые для исследованных систем металл – углеродная подложка приведены в таблице 1.

Параметры $\alpha$ и $\beta$ являются положительными, что свидетельствует об уменьшении поверхностной энергии микрочастиц и межфазной энергии на границе с подложкой с уменьшением радиуса. Значения $\alpha$ примерно соответствуют толщине переходного слоя на границе жидкой фазы с вакуумом. Величина $\beta$, характеризующая ширину переходной зоны между жидкой час-тицей и подложкой и зависящая от природы контактирующих фаз, в 2–4 раза больше α.

*Таблица 1*

### Результаты по размерному эффекту смачивания в системах металл – углерод [18]

| Металл | $\sigma_l^\infty$, мДж/м² | α, нм | | $\sigma_{ul}^\infty$, мДж/м² | β, нм | $\theta_\infty^\circ$ |
|:---:|:---:|:---:|:---:|:---:|:---:|:---:|
| | | расчет | экспер. [52] | | | |
| Au | 1130 | 0,48 | 0,55 | 970 | 2,2 | 138,8 |
| Sn | 531 | 0,52 | – | 582 | 1,0 | 152,7 |
| Pb | 450 | 0,58 | 0,64 | 474 | 2,1 | 141,8 |
| Bi | 376 | 0,60 | – | 415 | 1,3 | 141,4 |

Размерные эффекты при смачивании могут наблюдаться и в случае, когда высокодисперсной является твердая фаза, например, в системах типа жидкий металл – тонкая пленка – массивная подложка. Для систем с отсутствием химического взаимодействия, изменяя толщину промежуточной пленки от минимальной, обеспечивающей ее сплошность, до такой, когда пленку можно рассматривать как массивный материал, можно исследовать размерный эффект, связанный с изменением поверхностной энергии пленки с уменьшением ее толщины. Такие исследования были выполнены в работах [17, 28].

Для экспериментов [17, 28] была выбрана система «островковый жидкий конденсат олова – углеродная пленка – монокристалл KCl», в которой толщину углеродной пленки изменяли в пределах $2 < t < 30$ нм. Образцы препарировались путем испарения и конденсации углерода и олова в вакууме $4 \cdot 10^{-8}$ мм рт. ст. на подложки из монокристаллов KCl при температуре 520 K. Геометрия взаимного расположения испарителей и подложки выбиралась такой, которая обеспечивала в одном эксперименте получение образцов с существенно различной толщиной пленок углерода. Измерения углов проводились на закристаллизовавшихся каплях размером более 40 нм, чтобы исключить зависимость $\theta(R)$.

Было установлено, что при малых толщинах пленок углерода величина $\theta$ стремится к значению, соответствующему смачиванию чистой монокристальной подложки, т. е. $\theta(Sn/C/KCl) \to \theta(Sn/KCl)$ при $t \to 0$, а при больших $t$ угол соответствует смачиванию оловом материала пленки, т. е. $\theta(Sn/C/KCl) \to \theta(Sn/C)$ (рис. 15a). Интервал толщин пленок углерода, в котором происходит изменение $\theta$ в указанных пределах, составляет $2{,}5 < t < 7$ нм. Электронно-микроскопические исследования чистых пленок углерода указывают на их сплошность при уменьшении толщины вплоть до 1,5–2 нм.

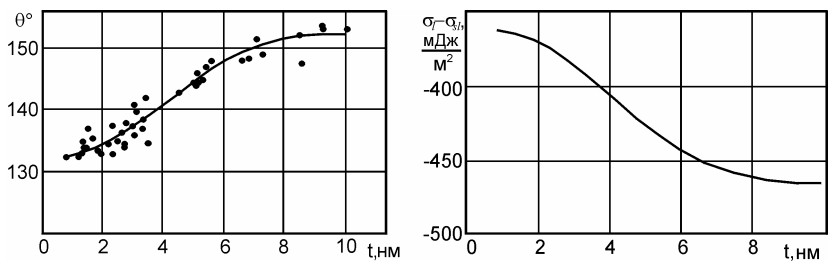

*Рис. 15. Зависимость краевого угла смачивания (а) и адгезионного натяжения (б) от толщины углеродной пленки для системы Sn/C/KCl [17, 28].*

Проведенные исследования [17, 28] позволили предположить, что зависимость $\theta(t)$, в соответствии с выражением (18), обусловлена изменением поверхностной энергии углеродных пленок с уменьшением их толщины. Так как при уменьшении $t$ изменяется не только поверхностная энергия пленки $\sigma_u$, но и межфазная энергия границы раздела пленка–частица $\sigma_{ul}$, то зависимость $\theta(t)$ в системе Sn/C/KCl, строго говоря, отражает изменение адгезионного натяжения $\sigma_u - \sigma_{ul}$ при уменьшении толщины углеродной пленки, находящейся на поверхности кристалла KCl. На рис. 15б приведена зависимость адгезионного натяжения от толщины пленок углерода для системы Sn/C/KCl, рассчитанная по данным $\theta(t)$. Увеличение адгезионного натяжения при уменьшении толщины углеродной пленки свидетельствует об уменьшении ее поверхностной энергии в предположении, что межфазная энергия границы тонкая пленка – капля уменьшается при уменьшении толщины, как это имеет место для границы подложка – микрочастица с уменьшением радиуса частицы.

## 2.3. Размерный эффект при смачивании упругодеформируемой подложки

Наблюдаемое уменьшение краевого угла $\theta$ с уменьшением радиуса капли объясняется размерной зависимостью $\sigma_l$ и $\sigma_{lu}$, вызванной возрастанием относительного вклада граничных областей, свойства вещества в которых существенно отличаются от объемных. Однако на параметры смачивания также оказывает влияние упругая деформация подложки силами поверхностного натяжения жидкости, которая не учитывалась выше при получении уравнений (30) и (31). Влияние деформации на угол $\theta$ в случае, когда подложкой является тонкая пленка, подробно рассмотрено в работе [56], а для упругого полупространства – в [31]. В силу использованного приближения полученные в работе [31] результаты неприменимы для капель размером менее 20–50 нм, т. е. когда наблюдается размерный эффект смачивания. В работе [32] решена задача об определении значения равновесного угла смачивания $\theta$ для микрокапли радиусом менее 50 нм с учетом упругой деформации подложки.

Рассмотрим каплю объемом *V*, помещенную на упругое изотропное полупространство [32]. Со стороны капли на тело действуют две силы: поверхностного натяжения и гидростатического давления. Предположим, что сила поверхностного натяжения жидкости равномерно распределена по кольцу с внешним радиусом $r$ и шириной $\tau$, где $r$ – радиус периметра смачивания, а $\tau$ имеет смысл эффективной толщины граничного слоя жидкости (рис. 16). Гидростатическое давление под каплей может быть записано следующим образом:

$$p = \frac{2\sigma_l}{\tau} \int\limits_{R-\tau}^{R} \frac{d\rho}{\rho} = -\frac{2\sigma_l}{\tau} \ln\left(1 - \frac{\tau}{R}\right); \qquad (34)$$

здесь $R = r/\sin\theta$ – радиус кривизны поверхности капли, $\rho$ – полярная координата, отсчитываемая от вертикальной оси капли.

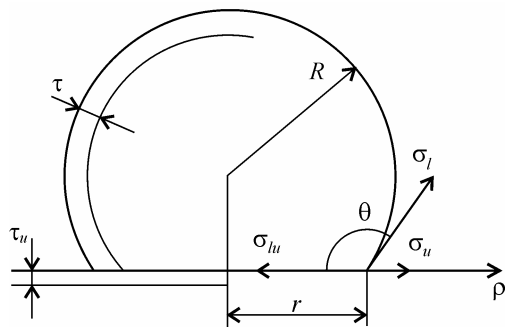

*Рис. 16. Схематическое изображение капли на подложке*

Давление в области периметра смачивания направленно по нормали к поверхности, его тангенциальной составляющей можно пренебречь, так как для равновесного значения краевого угла она полностью компенсируется силами поверхностного натяжения $\sigma_{lu}$ и $\sigma_u$, а при небольших отклонениях $\theta$ ее вклад, как показывают оценки, незначителен. Таким образом, давление со стороны капли на подложку можно записать в виде

$$p(\rho) = \begin{cases} -\dfrac{2\sigma_l}{\tau}\ln\!\left(1-\dfrac{\tau}{R}\right) & \rho \leq r-\tau; \\[2ex] \dfrac{2\sigma_l}{\tau}\dfrac{(r-\tau)^2}{2r\tau-\tau^2}\ln\!\left(1-\dfrac{\tau}{R}\right) & r-\tau < \rho \leq r; \\[2ex] 0 & \rho > r. \end{cases} \tag{35}$$

Упругая деформация рассчитывается по формуле из [81], которая в полярных координатах приобретает следующий вид:

$$U(\rho) = \frac{\left(1-\nu^2\right)}{\pi E}\int\limits_0^{2\pi}\int\limits_0^\infty \frac{p(\rho')}{\xi}\rho'd\rho'd\varphi, \tag{36}$$

где $\nu$ – коэффициент Пуассона, $E$ – модуль Юнга, $\xi$ – расстояние от точки $(\rho,\varphi)$ до элемента площади $d\rho'd\varphi$.

Соответствующая этой деформации упругая энергия определяется выражением

$$W = \pi\int\limits_0^r p(\rho)U(\rho)\rho d\rho. \tag{37}$$

Равновесные параметры системы могут быть определены путем минимизации функционала свободной энергии

$$F = -pV + \sigma_l S_l + \sigma_{lu}S_{lu} + \sigma_u S_u + W\,; \tag{38}$$

здесь $p$ – неопределенный множитель Лагранжа, учитывающий постоянство объема капли, $S_l,\ S_{lu},\ S_u$ – площади соответствующих поверхностей раздела. На искривленной поверхности их можно записать в виде

$$S_{lu} = 2\pi\int\limits_0^r\sqrt{1+\frac{\partial U(\rho)^2}{\partial\rho}}\,\rho d\rho, \quad S_u = 2\pi\int\limits_r^\infty\sqrt{1+\frac{\partial U(\rho)^2}{\partial\rho}}\,\rho d\rho. \tag{39}$$

Эффективная площадь поверхности жидкой фазы находится как $S_l = V_c/\tau$ ($V_c$ – объем поверхностного слоя жидкости), т. е.

$$S_l = 2\pi R^2 (1 - \cos\theta') \left[ 1 - \frac{\tau}{R} \left( 1 - \cos\theta' - \cos^2\theta' \right) \right], \qquad (40)$$

где $\theta'$ – краевой угол на деформированной подложке; его связь с углом смачивания на плоской поверхности следует из решения уравнения $V = V' + V''$, где $V' = \dfrac{\pi r^3}{3} \dfrac{(1 - \cos\theta')^2 (2 + \cos\theta')}{\sin^3\theta'}$ и

$V'' = 2\pi \displaystyle\int\limits_0^r [U_{\max} - U(\rho)] \rho\, d\rho$ – части объема выше и ниже уровня

периметра смачивания соответственно.

Можно отметить, что формула (40) определяет площадь гиббсовской поверхности натяжения. Соответствующий $S_l$ радиус кривизны определяется следующим выражением:

$$R' = R \left( 1 - \frac{\tau}{2R} \left( 1 - \cos\theta - \cos^2\theta \right) \right).$$

Коэффициент поверхностного натяжения жидкости, как известно, определяется избыточной энергией молекул граничного слоя. Для большой капли, в рамках предложенной модели, она равна

$$\Delta\mu_c^\infty = \frac{\sigma_l^\infty}{\tau} \omega^l,$$

где $\omega^l$ – атомный объем жидкости. С уменьшением капли гидростатическая энергия, приходящаяся на один атом, возрастает

$$\Delta\mu_v^{pv} = \omega^l p = \omega^l \left( -2\frac{\sigma_l^\infty}{\tau} \ln\left( 1 - \frac{\tau}{R} \right) \right).$$

Вклад гидростатической энергии для молекул поверхностного слоя – величина второго порядка малости по $\tau/R$, поэтому ею можно пренебречь. Таким образом, избыточная энергия атомов поверхностного слоя с удержанием членов порядка $\tau/R$

$$\Delta\mu_c = \Delta\mu_c^{\infty} - \Delta\mu_v^{pv} = \frac{\sigma_l^{\infty}}{\tau}\omega^l\left(1 - 2\frac{\tau}{R}\right),$$

а соответствующий ей коэффициент поверхностного натяжения

$$\sigma_l = \sigma_l^{\infty}\left(1 - 2\frac{\tau}{R}\right). \tag{41}$$

Аналогичным соотношением $\sigma_l = \sigma_l^{\infty}\left(1 - \delta/R\right)$ в работе [18] была аппроксимирована размерная зависимость $\sigma_l$ по экспериментальным данным о скорости испарения микрокапель. Используя результаты этой работы и очевидное соотношение

$$\tau = \delta/2, \tag{42}$$

можно найти толщину поверхностного слоя $\tau$, необходимую для расчета упругой деформации.

Размерная зависимость $\sigma_{lu}$ оценивается аналогичным образом. Введем эффективную толщину переходного слоя подложка – жидкость $\tau_u$. Энергия, приходящаяся на один атом граничного слоя, равна

$$\mu_c^u = \frac{\sigma_{lu}}{\tau_u}\omega^u + \mu_0^u;$$

здесь $\omega^u$ – атомный объем вещества подложки, а $\mu_0^u$ – химический потенциал ее атомов. Введем коэффициент $\eta$, характеризующий относительный вклад жидкой фазы в избыточную энергию межфазного слоя капля–подложка. Тогда на один атом приходится энергия

$$\Delta\mu_c^{u\infty} = \frac{\sigma_{ul}^{\infty}}{\tau_u}\omega^u + \mu_0^u - \frac{\mu_0^u + \eta\mu^l}{1+\eta} = \frac{\sigma_{ul}^{\infty}}{\tau_u}\omega^u + \frac{\eta}{1+\eta}\left(\mu_0^u - \mu^l\right),$$

где $\mu^l$ – химический потенциал атомов жидкости. Поверхностные эффекты изменяют потенциал твердой фазы на величину $\Delta\mu^u = \omega^u p$, а жидкой на

$$\Delta\mu^l = \begin{cases} \omega^l p & \rho \le r - \tau \\ \omega^l \dfrac{\sigma_l^\infty}{\tau} & r - \tau < \rho \le r \end{cases};$$

здесь $p$ определяется выражением (35); таким образом, учитывая воздействие капли, получим

$$\Delta\mu_c^u = \Delta\mu_c^{u\infty} + \frac{\eta}{1+\eta}\left(\omega^u p - \Delta\mu^l\right).$$

Рассмотрим отдельно области $\rho \le r - \tau$ и $r - \tau < \rho \le r$. В области $\rho \le r - \tau$ давление $p$ определяется выражением (34), и избыток энергии атомов поверхностного слоя равен

$$\Delta\mu_c^u = \Delta\mu_c^{u\infty} + \frac{\eta}{1+\eta}\, p\left(\omega^u - \omega^l\right),$$

а соответствующий коэффициент $\sigma_{lu}$ имеет вид

$$\sigma_{ul} = \sigma_{ul}^\infty + \tau_u \frac{\eta}{1+\eta}\frac{\sigma_l^\infty}{R}\left(1 + \frac{\tau}{R}\right)\left(1 - \frac{\omega^l}{\omega^u}\right). \tag{43}$$

Аналогичное рассмотрение области $r - \tau < \rho \le r$ приводит к следующему результату:

$$\sigma_{ul} = \sigma_{ul}^\infty - \rho\frac{\eta}{1+\eta}\left[\frac{\sigma_l^\infty}{R}\frac{(r-\tau)^2}{2r\tau - \tau^2}\left(1 + \frac{\tau}{R}\right) + \frac{\omega^l}{2\omega^u}\frac{\sigma_l^\infty}{\tau}\right]. \tag{44}$$

Объединяя выражения (43) и (44), окончательно получим размерную зависимость

$$\sigma_{ul} = \sigma_{ul}^\infty\left[1 - \frac{\tau_u}{R}\frac{2\eta}{1+\eta}\frac{\sigma_l^\infty}{\sigma_{ul}^\infty}\frac{\omega^l}{\omega^u}\left(1 + \sin\theta\right)\right]. \tag{45}$$

Выражения (37), (39)–(41), (45) определяют все величины, входящие в функционал свободной энергии (38). Но в формулы для поверхностных энергий входят параметры $\tau$ и $\tau_u$ – толщины переходных слоёв жидкость – пар и подложка – жидкость соответственно. С учётом соотношения (42) было использовано зна-чение $\tau = 0.24$ нм [18], а для определения коэффициента $2\tau_u\eta/(1 + \eta)$ проведено сопоставление величины угла θ, найденной путем минимизации функционала (38), с экспериментально наблюдаемым значением в точке $R = 10$ нм. Полученная таким образом толщина граничного слоя подложки, в предположении равного вклада фаз, составляет ~ 0.1 нм, что согласуется с известными представлениями о структуре поверхностного слоя. Рассчитанная на ЭВМ зависимость краевого угла смачивания от радиуса капли для системы Au/C представлена на рис. 17 (кривая 1), там же приведены экспериментальные значения θ по данным работ [18, 27], кривая 2, относительно пунктирной линии $\theta^\infty$, соответствует вкладу в эту зависимость энергии упругой деформации подложки.

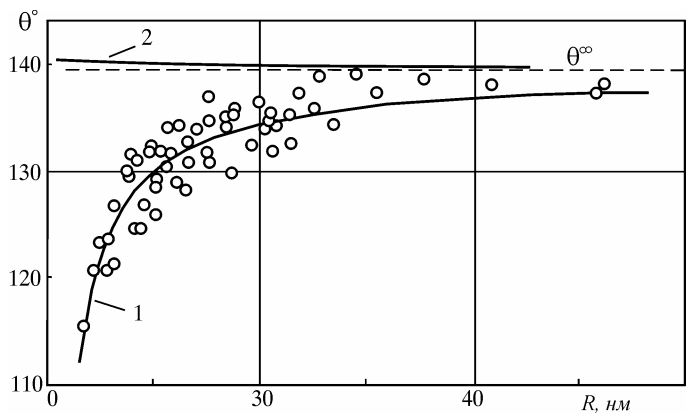

*Рис. 17. Зависимость краевого угла смачивания θ от радиуса капли R в системе Au/C [32]. Кривая 1 – равновесные значения θ, 2 – вклад упругой деформации, точки – экспериментальные данные [18, 27], пунктир – асимптотика для случая макроскопической капли*

Видно, что влияние упругой деформации незначительно, и уменьшение краевого угла в основном определяется размерной зависимостью удельных энергий поверхностей раздела. Однако, как следует из выражения (37), вклад упругой энергии про-порционален $1/E$, поэтому для подложек с малым модулем Юнга упругая деформация может оказаться существенной; оценки показывают, что для подложек с модулем Юнга $E \sim 10^9$ Н/м$^2$ дополнительное отклонение краевого угла смачивания для малых капель может достигать 5–6 градусов.

## 2.4. Гистерезис смачивания у конденсированных микрокапель

Как показано выше, размерная зависимость поверхностной энергии жидкой фазы $\sigma_l$ может быть обнаружена при измерениях краевого угла смачивания $\theta$, который связан с $\sigma_l$ уравнением Юнга. Также изменение $\sigma_l$ с размером капли обнаруживается при анализе результатов по ее испарению при постоянной температуре, скорость которого у достаточно малых капель пропорциональна $\exp(\sigma_l/r)$ [11, 18, 52]. При этом на кривой испарения, как правило, присутствуют периодические отклонения экспериментальных точек от плавной зависимости. В работах по непосредственному измерению краевого угла смачивания также наблюдаются колебания $\Delta\theta \sim 10$–$15°$, в то время как точность применяемых методов свертки и фотометрирования составляет 3–5° [12]. И в том, и в другом случае причиной таких отклонений может быть гистерезис смачивания. Суть этого явления состоит в фиксации периметра смачивания, что в некоторых условиях, например, при испарении, существенно меняет поведение жидкой капли. В работе [33] рассматриваются причины, обусловливающие этот эффект для микрокапель, и выполнена оценка влияния гистерезиса смачивания на параметры системы капля – подложка.

Изучению гистерезиса смачивания посвящен ряд работ [34, 35]; наиболее общими причинами этого эффекта считаются микронеровности и неоднородность подложки [35]. Однако многие допущения, положенные в основу этих работ, например, высота микронеровностей $\sim 1{,}25$ мкм, неприменимы к микрокаплям. В некоторых системах [35] фиксация периметра смачивания достигается частичным взаимным растворением твердой и жидкой фаз. Тем не менее, гистерезис может наблюдаться и для систем с пренебрежимо малой взаимной растворимостью, таких как Au/C. Некоторыми авторами [34] отмечалось, что при высоких температурах под действием сил поверхностного натяжения жидкости подложка может пластически деформироваться. При этом в зоне тройного контакта образуется выпуклый рант. Сравнение различных механизмов массопереноса на малых ($\sim 10^{-8}$ м) расстояниях [36] позволяет сделать вывод об определяющей роли поверхностной диффузии. В области тройного контакта сила поверхностного натяжения жидкости вызывает локальное снижение химического потенциала $\Delta\mu_P$, соответствующее приложенному давлению $\Delta\mu_P = -\sigma_l/\tau$. Известно также, что химический потенциал атомов на искривленной поверхности имеет добавку $\Delta\mu_C = \sigma_{lu}\,C$ ($\sigma_{lu}$ – удельная поверхностная энергия границы раздела подложка – жидкость, $C$ – кривизна). Таким образом, энергетически выгодное выравнивание градиента химического потенциала подложки может осуществляться путем ее искривления. Оценим характерное время такой деформации. Как следует из работы [36], скорость деформирования равна

$$\frac{dU}{dt} = -\frac{D_S \omega^S n_0}{kT}\frac{\partial^2 \mu}{\partial S^2}\,, \qquad (46)$$

где $D_S$ – коэффициент поверхностной диффузии, $n_0$ – поверхностная концентрация атомов подложки, $\omega^S$ – их объем. Изменение химического потенциала $\mu$ на поверхности $S$ происходит, главным образом, в области периметра смачивания на расстоянии порядка ширины зоны тройного контакта, которую в пер-вом приближении можно считать равной толщине переходного слоя жидкости τ.

Выравнивание химического потенциала $\Delta\mu = \Delta\mu_P + \Delta\mu_C = 0$ в идеализированном случае достигается при кривизне $C = \dfrac{\sigma_l}{\sigma_{lu}}\dfrac{1}{\tau}$. Необходимо отметить, что в области тройного контакта величина $\sigma_{lu}$ должна отличаться от своего макроскопического значения и, по всей видимости, составляет нечто среднее между $\sigma_{lu}$ и $\sigma_u$. Высота ранта $h$ с кривизной $C$, к примеру, для микрокапли золота на углеродной подложке составит тогда $h \sim 0{,}1$–$0{,}2$ нм. Принимая $D_S \sim 10^{-15}$ м²/с, $kT \sim 1{,}6{\cdot}10^{-20}$ Дж, $n_0 \sim 1{,}5{\cdot}10^{19}$ м⁻², $\omega^S \sim 1{,}7{\cdot}10^{-30}$ м³, $\sigma_l \sim 1$ Дж/м², $\tau \sim 10^{-9}$ м, получим характерное время деформирования:

$$t_C = h\,\frac{kT}{D_S n_0 \omega^{S^2}}\,\frac{\tau^3}{\sigma_l} \approx 10^{-1} c\ .$$

Таким образом, принимая во внимание время эксперимента (обычно $\sim 10^2$–$10^3$ с), можно сделать вывод, что рант успевает образовываться даже при довольно частых скачках периметра смачивания. То есть можно сказать, что капля сама создает микронеровности подложки. В качестве примера на рис. 18а показан сектор подложки, рассчитанный как решение дифференциального уравнения (46); для наглядности масштаб по вертикальной оси несколько увеличен. При рассмотрении гистерезиса смачивания будем считать, что профиль подложки имеет вид рис. 18б; из сравнения рис. 18а и рис. 18б видна правомерность такой аппроксимации.

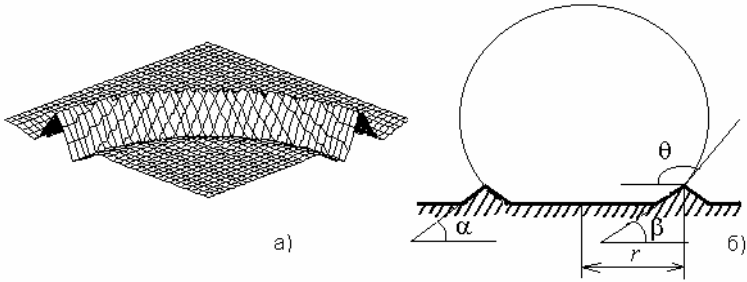

*Рис. 18. Профиль подложки после деформации: а) полученный как реше-
ние уравнения (46), б) принятый в работе для анализа гистерезиса [33]*

Энергетически выгодное положение определяется мини-
мумом функционала свободной энергии $F$

$$F = \sigma_l S_l + (\sigma_u - \sigma_{ul}) S_{ul} + W - pV + \mu N ;$$

здесь $W$ – энергия упругой деформации подложки; $S_l$, $S_{ul}$ –
площади поверхностей раздела соответствующих фаз. В при-
ближении $r \gg \tau$, то есть когда толщина переходного слоя гораз-
до меньше радиуса периметра смачивания $r$, указанные площа-
ди определяются следующими соотношениями: $S_l = \dfrac{2\pi r^2}{1 + \cos\theta}$ ;

$S_{ul} = \pi r^2$ . Необходимо отметить, что в области периметра сма-
чивания ее нормальная составляющая $W_n$ практически полно-
стью релаксирует при пластической деформации. Величина $W_n$
определена в работе [31]

$$W_n = \frac{3}{4} \frac{\sigma_l^2 \, r \sin^2\theta}{\pi E} \left( \ln 2 + \ln\frac{r}{\tau} - \frac{1}{2} \right).$$

Рассмотрим изменение свободной энергии капли $\delta F$ при из-
менении ее объема на $\delta V$ в двух случаях: при постоянном краевом
угле $\theta$ и при фиксированном периметре смачивания $r$. Очевидно,
что будет реализовано энергетически более выгодное состояние.

$$\delta F\big|_{\theta=\text{const}} = \left[\sigma_l \frac{\partial S_l}{\partial a} + \left(\sigma_u - \sigma_{ul}\right)\frac{\partial S_{ul}}{\partial a} + \frac{\partial W}{\partial a}\right]\frac{\partial a}{\partial V}\delta V - p\delta V + \mu\delta N\,;$$

$$\delta F\big|_{r=\text{const}} = \sigma_l \frac{\partial S_l}{\partial \cos\theta} \cdot \frac{\partial \cos\theta}{\partial V}\delta V - p\delta V + \mu\delta N\,.$$

Частные производные площадей поверхностей раздела на подложке, изображенной на рис. 18a, определяются дифференцированием

$$dS_l = -\frac{2\pi r^2}{(1+\cos\theta)^2}\,d\cos\theta + 2\pi r\left[\frac{2}{1+\cos\theta} + \frac{\sin\gamma}{\sin(\theta-\gamma)}\right]dr\,;$$

$$dS_{ul} = 2\pi r \frac{\sin\theta}{\sin(\theta-\gamma)}dr\,;$$

где $\gamma = \begin{cases} -\beta; & \delta V < 0 \\ \;\;\alpha; & \delta V > 0 \end{cases}$ .

Для упругой энергии следует выражение

$$dW = W_n \frac{2|dr|}{\tau} = 2\pi r\sigma_l\chi|dr|, \qquad (47)$$

где $\chi = \dfrac{3}{4}\dfrac{\sigma_l \sin^2\theta}{\pi^2 E\tau}\left(\ln 2 + \ln\dfrac{r}{\tau} - \dfrac{1}{2}\right).$

Можно показать, что остальные слагаемые, входящие в $dW$, имеют порядок малости $\tau/r$ и потому не учитываются. Производные $\dfrac{\partial r}{\partial V}$ и $\dfrac{\partial \cos\theta}{\partial V}$ находятся из выражения для объема капли $V = \dfrac{\pi r^3}{3}\dfrac{(1-\cos\theta)^2(2+\cos\theta)}{\sin^3\theta}$;

$$\frac{\partial r}{\partial V} = \frac{1}{\pi r^2}\frac{\sin^3\theta}{(1-\cos\theta)^2(2+\cos\theta)}\,; \quad \frac{\partial \cos\theta}{\partial V} = -\frac{1}{\pi r^3}(1-\cos\theta)^2\sin\theta\,.$$

Значения краевых углов, при которых еще не наблюдается срыв периметра смачивания, могут быть получены из условия его фиксации $\delta F\big|_{\theta=\text{const}} > \delta F\big|_{a=\text{const}}$ в следующем виде:

$$\cos\theta_0 \, \frac{\sin\theta}{\sin(\theta+\beta)} + \frac{\sin\beta}{\sin(\theta+\beta)} + \chi < \cos\theta \qquad \delta V < 0 \, ; \qquad (48)$$

$$\cos\theta_0 \, \frac{\sin\theta}{\sin(\theta-\alpha)} - \frac{\sin\alpha}{\sin(\theta-\alpha)} - \chi < \cos\theta \qquad \delta V > 0 \, ; \qquad (49)$$

здесь $\theta_0$ – равновесный угол смачивания в соответствии с уравнением Юнга, для малых капель следует учитывать его размерную зависимость [11, 18].

Критические углы, соответствующие уравнениям (48) и (49), называются, соответственно, углами оттекания $\theta_r$ и натекания $\theta_a$. Если не учитывать упругую энергию деформации, они принимают очевидные значения: $\theta_a = \theta_0 + \alpha$; $\theta_r = \theta_0 - \beta$. Стоит отметить, что вклад релаксировавшей энергии упругой деформации достигает заметных значений, например, для системы Au/C ($\sigma_l \sim 1$ Дж/м$^2$, $E \sim 4{\cdot}10^{10}$ Н/м$^2$, $R \sim 10$ нм) $\chi \sim 0{,}025$ и при $\theta \sim 120°$ соответствующая разность $\theta_a - \theta_r \sim 3°$.

Таким образом, при изменении объема капли, например, при испарении, ее периметр смачивания будет неподвижным до тех пор, пока выполняется условие (48), при этом краевой угол смачивания будет плавно уменьшаться до критического значения $\theta_r$. Затем произойдет срыв периметра смачивания и капля примет положение, соответствующее юнговскому значению $\theta_0$. Довольно быстро на новом положении периметра смачивания образуется новый рант, и процесс повторится. Можно заметить, что при выполнении условия $r \gg \tau$ величина углов $\theta_r$ и $\theta_a$ слабо зависит от радиуса капли. Так, для системы Au/C при его увеличении от 20 до 1000 нм величина этих углов изменяется на 1° за счет увеличения вклада упругой энергии. Иллюстрирует гистерезис смачивания рис. 19, на котором показана кривая испарения капли Au, численно рассчитанная как решение уравнения для испаряющейся частицы с учетом меняющегося краевого угла и скачков пери-метра смачивания. Точками на рисунке отмечены эксперимен-
тальные данные [37].

Значительная величина лапласова давления у очень малых
(менее 10 нм) капель приводит к тому, что упругая деформация
успевает релаксировать не только в области тройного контакта, но
и непосредственно под каплей. В этом случае профиль подложки
будет иметь несколько иной вид: наряду с рантом по периметру
возможно образование лунки под каплей, что приводит к увели-
чению величины гистерезиса, при этом разность $\theta_0 - \theta_r$ увеличится
значительно сильнее, чем $\theta_a - \theta_0$. Принимая во внимание размер-
ное уменьшение краевого угла, можно сделать вывод, что, напри-
мер, для той же системы Au/C краевой угол малых ($\sim$2–5 нм) ка-
пель может достигать 65–70° при $\theta_0 = 138°$. Подтверждением это-
го заключения могут служить полученные из анализа рис. 19 раз-
мерные зависимости $\theta - \theta_r$ и $\theta_0 - \theta_r$ (рис. 20).

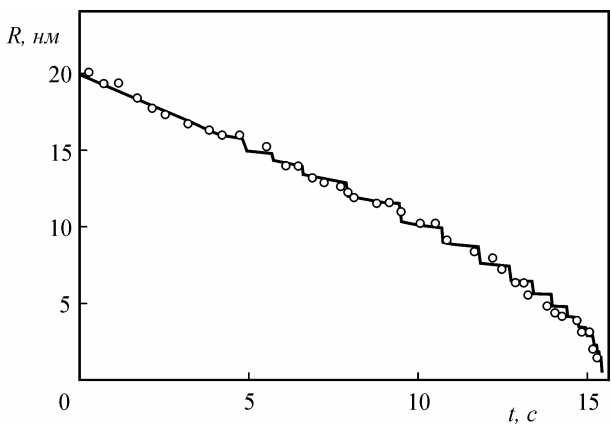

*Рис. 19. Кривая испарения капли Au с учетом гистерезиса смачивания,
точки – экспериментальные данные [37]*

Как видно из рис. 20б, разность $\theta - \theta_r$ для малых капель дос-
тигает величины $\sim$ 16°, благодаря чему на подложке могут устой-
чиво существовать капли, к примеру, образовавшиеся в результа-
те слияния, периметр смачивания которых не будет иметь форму окружности вследствие того, что капля может частично оставаться на рантах, образовавшихся до соприкосновения капель.

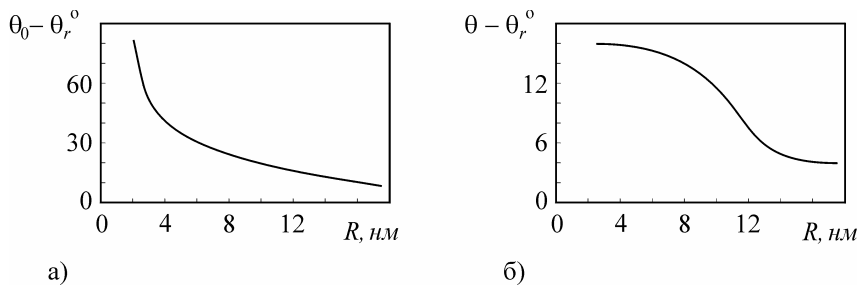

*Рис. 20. Размерная зависимость отклонения угла оттекания от равновесного значения $\theta_0$ для массивных капель (а) и с учетом размерного эффекта смачивания (б) в системе Au/C [32]*

Необходимо отметить, что, как видно из выражений (47, 48, 49), вклад энергии упругой деформации и высота ранта пропорциональны поверхностному натяжению капли $\sigma_l$, из чего следует, что для капель с малым $\sigma_l$ величина гистерезиса будет значительно меньше. Однако в реальных условиях на подложке практически неизбежно присутствуют адатомы посторонних веществ, которым энергетически выгодно, попадая в зону тройного контакта капли с подложкой, оставаться там. Таким образом, при наблюдении микрокапель гистерезис смачивания может быть обнаружен и в таких системах; соответственно, при анализе их поведения также необходимо учитывать возможные отклонения краевого угла смачивания от юнговского значения.

*Глава 3*

## Смачивание в системах «островковый конденсат – тонкая пленка – подложка»

Исследования смачивания в ультрадисперсных системах металл – металл в зависимости от степени дисперсности жидкой и твердой фаз представляют научный и практический интерес, поскольку дают возможность изучать влияние геометрических и физико-химических параметров на контактное взаимодействие и позволяют целенаправленно управлять смачиванием в различных системах при решении технологических задач.

Среди факторов, определяющих смачивание в дисперсных системах металл-металл, кроме размерных эффектов, связанных с зависимостью поверхностной и межфазной энергии от размеров фаз, можно выделить следующие [15, 17]: несплошность промежуточной пленки и связанная с ней гетерогенность подложки; растворимость компонентов друг в друге; образование химических соединений на границе твердой и жидкой фаз; окисление металлической пленки, степень которого будет определяться физико-технологическими параметрами процесса препарирования образцов. Для поликристаллических пленок следует ожидать зависимости смачивания от соотношения характерных размеров жидких капель и кристаллических зерен в пленке-подложке вследствие отличия поверхностной энергии для различных граней зерен и др. Таким образом, процессы смачивания в ультрадисперсных системах определяются целым рядом параметров, разделить которые достаточно сложно.

Примером исследования влияния дисперсности твердой фазы на краевой угол явились работы, выполненные коллективом авторов [38, 39, 40], в которых изучалось смачивание тонких пленок переменной толщины, нанесенных на массивную подложку. В работах [38, 39] было показано, что в системе «расплав (Ag, Cu, Sn, Pb) – металлическая пленка (Mo, V, Fe) – неметаллическая подложка (сапфир, кварц, графит)» краевой угол линейно изменяется с толщиной пленки в пределах значений, соответствующих смачиванию неметаллической подложки (при толщинах пленки $t \to 0$) и смачиванию вещества пленки в компактном состоянии (при $t > t_k$). Значения критических толщин $t_k$, ниже которых наблюдается изменение краевого угла, для исследованных систем находятся в пределах 20–50 нм.

В работе [40] исследовалось смачивание оловом неметаллических пленок, нанесенных на массивные металлические подложки. Экспериментальные данные показали, что смачивание германия, покрытого углеродом, резко ухудшается с ростом толщины пленок, а начиная с $t \geq 3$ нм остается неизменным и соответствует смачиванию компактного графита. Полученные результаты объясняются несплошностью пленок углерода при $t < 3$ нм. В системе $Sn/Al_2O_3/Mo$, согласно данным работы [40], изменение краевого угла начинается при толщинах пленок $Al_2O_3$ $t \approx 80$ нм, что, по мнению авторов, объясняется физико-химическим взаимодействием $Al_2O_3$ с молибденом при высоких температурах (1100 K).

Описанные эксперименты по смачиванию в трехкомпонентных системах [38, 39, 40] не позволили установить характер изменения поверхностной энергии пленки с ее толщиной ввиду сложности процессов взаимодействия расплава с дис-персной твердой фазой. Однако результаты этих исследований открывают пути к целенаправленному управлению смачиванием, что может быть использовано на практике.

В работах [17, 28, 41, 42, 43] проведены исследования смачивания в тройных системах Pb/Ni/[NaCl, Si, GaAs], Sn/[C, Al, $Al_2O_3$]/KCl, Bi/Fe/KCl в зависимости от толщины металлической пленки (2 нм < $t$ < 200 нм) и размера жидких островков (5 нм < $R$ < $10^4$ нм). Выбор таких объектов и пределов изменения параметров $t$ и $R$ обусловлен тем, что указанные системы существенно различаются характером взаимодействия: Sn–C, Sn–$Al_2O_3$, Bi–Fe – полная нерастворимость в твердом и жидком состояниях, Sn–Al – растворимость 0,5 вес. % Al в Sn и Pb–Ni – растворимость до 4 вес. % Ni в Pb [44].

Образцы для исследований [17, 28, 41, 42, 43] препарировались следующим образом. На монокристальные подложки (KCl, NaCl, Si, GaAs) в вакууме $10^{-6}$–$10^{-8}$ мм рт. ст. конденсировалась промежуточная пленка переменной толщины (Al, Fe, Ni, C, $Al_2O_3$), на которую без нарушения вакуума по механизму пар→жидкость конденсировался исследуемый металл (Sn, Bi, Pb). Температура подложки во время конденсации была 653 K для Pb, 523 K для Sn и 560 K для Bi. Взаимное расположение испарителей и подложки позволяло в одном эксперименте получать серию образцов с различной толщиной промежуточной пленки. После охлаждения в вакууме пленки исследовались с помощью электронной и оптической микроскопии. Краевые углы смачивания (θ) измерялись на закристаллизовавшихся каплях с помощью изложенных выше (глава 1) специально разработанных методик [9, 11, 12]. При этом установлено, что угол смачивания зависит от толщины пленок, а в некоторых случаях и от размера островков при постоянной толщине.

Во всех исследованных системах степень смачивания существенно зависит от толщины пленок, но интервал толщин, в котором происходит изменение θ, оказывается различным, при этом зависимость θ($t$) для разных систем обусловлена различными механизмами. Общим для исследованных систем является то, что краевой угол в первом приближении определяется гетерогенностью смачиваемой поверхности и изменяется в крайних пределах, соответствующих смачиванию чистой подложки ($t \to 0$) и материала пленки в массивном состоянии ($t > t_k$). Критическая толщина $t_k$, при которой наблюдается полное экранирование массивной подложки тонкой пленкой, зависит от характера взаимодействия компонентов системы и изменяется от единиц нанометров (взаимодействие отсутствует) до десятков и сотен (растворение пленки в расплаве, образование химических соединений).

Наименьшие значения критической толщины обнаружены у систем с невзаимодействующими компонентами [17, 28] $Sn/C/KCl$ (рис. 15а, 21) и $Sn/Al_2O_3/KCl$ (рис. 21). Поскольку изменение смачивания в таких системах наблюдается при толщинах пленок менее 10 нм, в данном случае, как отмечено выше (раздел 2.2), возможно также проявление эффектов, связанных с изменением поверхностной энергии пленки с уменьшением ее толщины. Так в системе $Sn/C/KCl$ изменение θ наблюдается в интервале $2 < t < 7$ нм. В то же время электронно-микроскопические исследования чистых пленок углерода указывают на их сплошность при уменьшении толщины вплоть до 1,5–2 нм. Это дает основания предположить, что зависимость θ($t$) в системе $Sn/C/KCl$ обусловлена, кроме несплошности углеродных пленок, еще и изменением их поверхностной энергии. Так как при уменьшении $t$ изменяется не только поверхностная энергия пленки $\sigma_u$, но и межфазные энергии границ раздела пленка – частица $\sigma_{lu}$ и пленка – массивная подложка, то указанная зависимость, строго говоря, отражает изменение адгезионного на-тяжения $\sigma_u - \sigma_{lu}$ при уменьшении толщины углеродной пленки, находящейся на поверхности макроскопического монокристалла KCl (рис. 15б).

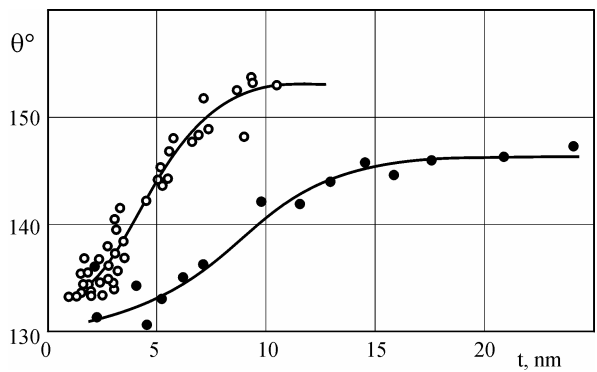

*Рис. 21. Краевой угол смачивания оловом пленок углерода (○) и Al₂O₃ (●)*
*различной толщины, нанесенных на поверхность KCl*

Наличие растворения материала промежуточной пленки в жидких каплях вызывает заметное смещение $t_k$ в область больших значений толщины; примеры изменения краевого угла смачивания в таких системах приведены на рис. 22.

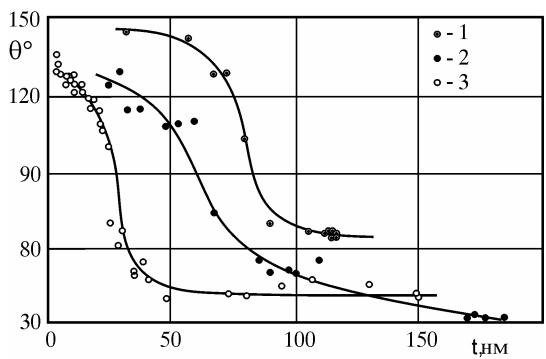

*Рис. 22. Зависимость θ(t) для систем*
*Bi/Fe/KCl (1), Sn/Al/KCl (2), Pb/Ni/NaCl (3) [17]*

В системе Pb/Ni/NaCl в интервале толщин пленок никеля 5–40 нм происходит изменение θ в пределах θ → θ(Pb/NaCl) при

$t < 5$ нм и $\theta \rightarrow \theta(\text{Pb/Ni})$ при $t > 40$ нм, т. е. при малых толщинах пленок никеля величины $\theta$ стремятся к значению, соответствующему смачиванию чистой подложки NaCl, а при больших $t$ угол $\theta$ соответствует смачиванию жидким металлом массивного никеля. Электронно-микроскопические и электронографические исследования показывают, что зависимость $\theta(t)$ в основном обусловлена растворением никеля в свинце, которое приводит к несплошности пленок никеля; т. е. при некоторых толщинах пленок подложка становится гетерогенной. Поскольку растворимость в системе Pb–Ni ограничена, то степень гетерогенности подложки будет зависеть от толщины пленки, что и обусловливает наблюдаемую зависимость $\theta(t)$.

В системе Pb/Ni/Si изменение $\theta$ с толщиной пленки происходит, в основном, в интервале $10 < t < 40$ нм, однако, в отличие от системы Pb/Ni/NaCl, при толщинах пленок никеля 30–80 нм наблюдается значительный разброс значений углов смачивания как для отдельных капель свинца, так и для усредненных по образцу величин (рис. 23).

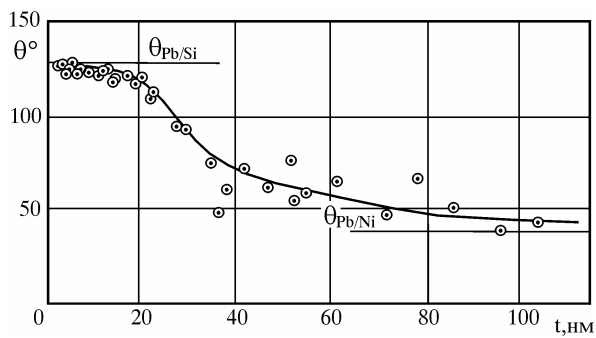

*Рис. 23. Краевые углы смачивания островковыми конденсатами свинца пленок никеля переменной толщины на монокристаллах кремния [42]*

Специально проведенными электронно-микроскопическими и электронографическими исследованиями показано [28], что наблюдаемая зависимость $\theta(t)$ в системе Pb/Ni/Si обуслов-лена образованием при конденсации дисилицида никеля, который лучше смачивается свинцом, чем чистый кремний. При увеличении толщины пленки никеля на подложке остается слой непровзаимодействовавшего металла, что приводит к уменьшению краевого угла смачивания до значения θ(Pb/Ni).

В системе Pb/Ni/GaAs [43], как и в других изученных тройных системах с промежуточным слоем [17, 28, 38, 39, 40, 41, 42], угол смачивания изменяется в крайних пределах, соответствующих смачиванию свинцом чистого арсенида галлия ($t \to 0$, θ ≈ 120°) и компактного никеля ($t < 20$ нм, θ ≈ 20°). На рис. 25 представлены микроснимки частиц свинца на Ni/GaAs подложке, иллюстрирующие изменение в морфологической структуре пленок. Увеличение толщины подслоя приводит к укрупнению островков Pb (при неизменной массовой толщине пленки свинца) и к переходу от сферических частиц, наблюдающихся обычно в случае плохого смачивания, к лабиринтным образованиям, характерным для растекания жидкой фазы по поверхности с хорошим смачиванием.

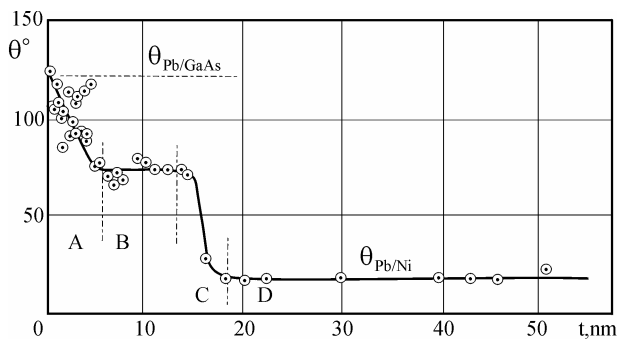

*Рис. 24. Зависимость краевого угла смачивания от толщины пленки никеля в системе Pb/Ni/GaAs [43]*

Как видно из рис. 24, в системе Pb/Ni/GaAs изменение смачивания происходит в два этапа. Сначала краевой угол уменьшается до промежуточного значения θ ≈ 75°, затем в интервале толщин $6 < t < 14$ нм на зависимости $\theta(t)$ наблюдается плато, которое не обнаруживалось в ранее изученных аналогичных системах, и далее снова происходит изменение краевого угла до значения, соответствующего системе Pb/Ni. Такой ход зависимости $\theta(t)$ прямо указывает на существование по крайней мере двух механизмов изменения смачивания с толщиной пленки, один из которых реализуется в интервале $0 < t < 14$ нм, а другой при $t > 14$ нм.

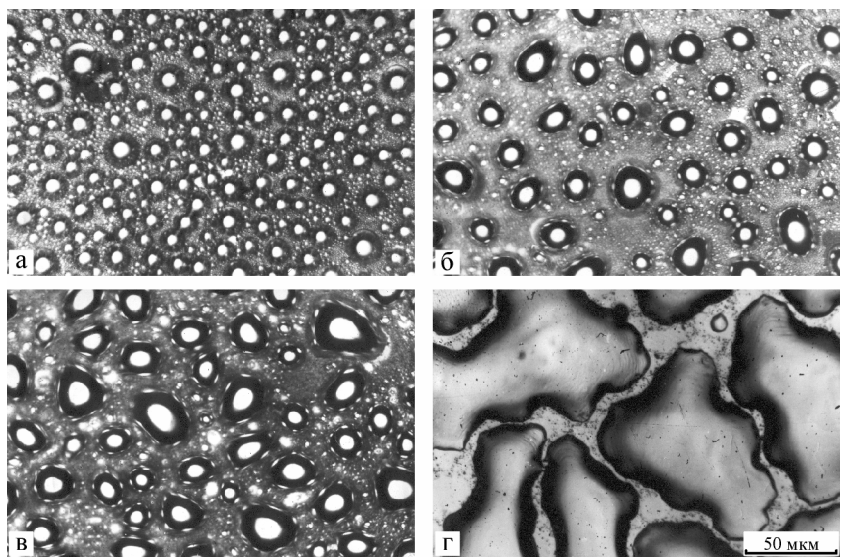

*Рис. 25. Микроснимки частиц свинца на пленках никеля различной*
*толщины (а – 1 нм; б – 4 нм; в – 10 нм; г – 30 нм), нанесенных на GaAs*

Сопоставление полученных данных с результатами исследований аналогичных систем Pb/Ni/Si и Pb/Ni/NaCl [9, 17, 42] позволяет сделать вывод, что изменение краевого угла в системе Pb/Ni/GaAs связано с образованием хорошо смачиваемых соединений на границе пленка никеля – монокристалл GaAs и с растворением никеля в жидком свинце. Исходя из этого, зависимость $\theta(t)$ можно разбить на четыре области, в которых реализуются различные состояния смачиваемой поверхности: 0–6 нм (A), 6–14 нм (B), 14–18 нм (C) и $t > 18$ нм (D) (рис. 24). Две области являются переходными, и смачивание в них происходит на поверхности с различной степенью гетерогенности, которая реализуется разными способами: путем образования химических соединений (область A) и путем растворения никеля в жидком свинце (область C).

Таким образом, механизм изменения краевого угла в системе Pb/Ni/GaAs с ростом толщины пленки никеля состоит в следующем. При $t \to 0$ краевой угол соответствует смачиванию чистого арсенида галлия: $\theta \approx 120°$. Конденсация на поверхности GaAs никеля приводит к образованию островков химического соединения, которое лучше смачивается свинцом, чем чистый GaAs. В области A с увеличением количества осажденного никеля возрастает доля поверхности, заполненной этим соединением, что приводит к плавному уменьшению краевого угла до значения $\theta \approx 75°$.

Далее пленка соединения становится сплошной, т. е. полностью экранирует подложку, и изменение смачивания прекращается (область B). Однако за время эксперимента в реакцию с подложкой может вступить ограниченное количество никеля. Поэтому, начиная с некоторой толщины $t < 14$ нм, на подложке остается слой непрореагировавшего никеля, который до $t \sim 14$ нм растворяется в свинце полностью, а при $t > 14$ нм часть его сохраняется на подложке, что приводит к гетерогенности смачиваемой поверхности и, следовательно, к появлению переходной области C. При $t > 18$ нм (область D) пленка никеля становится сплошной, а краевой угол смачивания – постоянным: $\theta \approx 20°$.

Для выяснения характера взаимодействия пленки никеля с подложкой были выполнены электронографические исследования двухслойных пленок Ni/GaAs, полученных вакуумной кон-денсацией на NaCl – подложке при температуре 670 К (для GaAs применялся метод дискретного испарения). Результаты расшифровки электронограмм, полученных от образцов с различной толщиной никелевой пленки, следующие. При $t > 20$ нм на электронограммах присутствуют только яркие линии никеля; рефлексы от других фаз не обнаруживаются, вероятно, ввиду их малой интенсивности. Начиная с толщин $t < 18$ нм, появляются отражения от фаз γ-$Ni_3Ga_2$ (гексагональная решетка типа NiAs) и NiAs. С уменьшением толщины пленки никеля ($t < 10$ нм) интенсивность рефлексов от этих фаз увеличивается и обнаруживаются также слабые дополнительные линии, которые могут принадлежать фазам γ-$NiAs_2$, $Ni_5As_2$, $Ni_2Ga$, известным из диаграмм равновесия двойных систем Ni–Ga и Ni–As [44]. Подробно результаты фазового анализа двухслойных пленок Ni–GaAs, полученных в различных условиях, приведены в работах [45–50]. Важно отметить, что в образцах Ni/GaAs при $t \sim 12$ нм электронографически не обнаруживается наличие чистого никеля, что подтверждает выводы, сделанные на основе анализа зависимости θ($t$) в системе Pb/Ni/GaAs.

Для системы Bi/Fe/KCl, которая характеризуется полной нерастворимостью компонентов в твердом и жидком состояниях, также наблюдается зависимость θ($t$), но при значительно больших толщинах пленок Fe, чем для системы Pb/Ni/NaCl. Из данных, представленных на рис. 22, следует, что при толщинах пленки железа $t < 60$ нм значение угла θ = 145°, т. е. существенно превышает величину для системы Bi/KCl (θ = 120°). Это означает, что верхний предел смачивания в системе Bi/Fe/KCl нельзя объяснить несплошностью пленок железа. В соответствии с электронно-микроскопическими данными и результатами качественного электронографического анализа зависимость θ($t$) в данной системе вызвана окислением пленок железа, которое может происходить по механизму окисления межфазной границы металл – щелочногаллоидный кристалл [26], а также, вероятно, за счет поступления кислорода из остаточной атмосферы вакуумной камеры. В связи с этим более тонкие пленки могут иметь слой окисла большей толщины, и, следовательно, для них межфазное взаимодействие на границе капля – подложка будет более ослаблено. Это должно приводить к наблюдаемому возрастанию угла смачивания при уменьшении толщины пленок железа в системе Bi/Fe/KCl.

Система Sn/Al/KCl характеризуется слабой растворимостью алюминия в олове и в этом смысле является промежуточной среди рассмотренных выше. В результате измерений [41], проведенных на препарированных в вакууме $10^{-7}$–$10^{-8}$ мм рт. ст. образцах, установлено, что в данной системе угол смачивания зависит как от толщины пленки алюминия, так и от размера частиц олова при постоянной толщине пленок алюминия.

При фиксированной толщине пленок алюминия угол смачивания $\theta$ растет с уменьшением размера частиц олова. На рис. 26а представлены результаты измерения $\theta$ для пленки алюминия толщиной 110 нм, а на рис. 26б – для пленки толщиной 40 нм в зависимости от радиуса капель олова. В первом случае угол смачивания увеличивается от 50° (для капель размеров порядка нескольких микрометров) до 110° (для капель радиусом в десятки нанометров). При толщине пленки алюминия 40 нм капли радиусом 25–40 нм имеют примерно такой же угол смачивания $\theta \approx 110°$. Для частиц олова, радиус которых менее 20 нм, угол $\theta$ возрастает до 140° с уменьшением размера капель, в отличие от ранее установленного уменьшения угла $\theta$ с уменьшением размера частиц (см. раздел 2.2), т. е. в данном случае наблюдается аномальный размерный эффект. Это явление не связано с размерным эффектом смачивания, а объясняется [41] тем, что более крупные капли конденсировались ранее мелких, и подслой окиси алюминия, успевший образоваться до конденсации олова, под ними более тонкий, чем под мелкими. Увеличение толщины подслоя окиси алюминия приводит к ослаблению межфазного взаимодействия, т. е. к возрастанию угла смачивания.

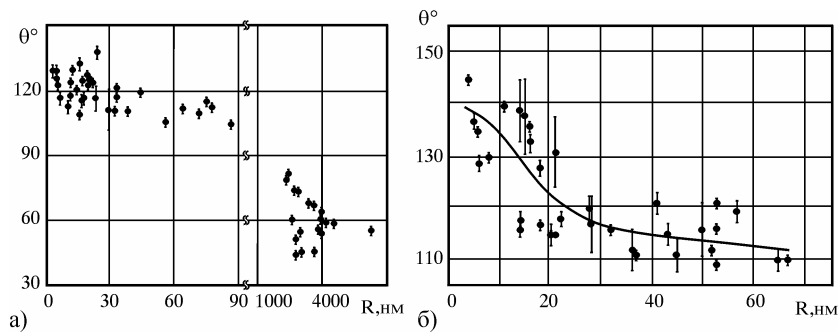

*Рис. 26. Зависимость θ(R) в системе Sn/Al/KCl*
*при толщине пленок алюминия 110 (а) и 40 (б) нм [17, 41]*

Детальные электронно-микроскопические наблюдения [41] показывают, что на начальных стадиях конденсации олово осаждается по границам зерен алюминия, смачивает его и при дальнейшей конденсации обволакивает зерна алюминия. Сферические капли олова на зернах – это осажденные позднее порции олова, сконденсировавшиеся на уже успевшую окислиться поверхность зерен алюминия. По мере увеличения количества олова зерна алюминия приобретают в результате растворения и перехода в жидкую фазу характерную «желудеобразную» форму. Под крупными растекшимися каплями олова, как правило, сохраняется некоторый подслой алюминия. Аморфная пленка $Al_2O_3$ образуется на границе кристалла KCl и пленки алюминия. Кроме того, поверхность пленки алюминия окисляется за счет остаточных газов. Следовательно, дальнейшая конденсация происходит на более толстом слое окиси алюминия, поэтому самые маленькие частицы олова имеют угол смачивания примерно 140°, что соответствует, по проведенным измерениям, углу смачивания оловом пленки окиси алюминия (см. рис. 21).

В работе [41] изучено также влияние толщины слоя алюминия на угол смачивания. На рис. 27а представлена зависимость угла смачивания от толщины пленки алюминия, а на рис. 27б – от размера частиц олова для примерно одинаковых соотношений между количествами сконденсированных олова и алюминия по данным ряда экспериментов.

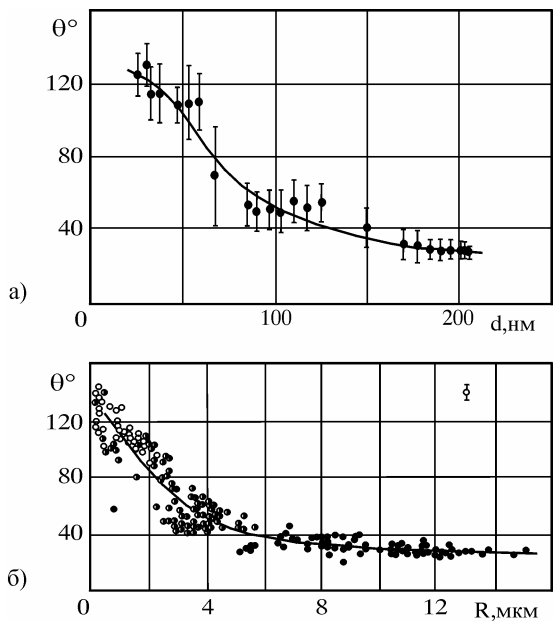

*Рис. 27. Зависимость угла смачивания каплями олова пленок алюминия на KCl подложке [41] от их толщины (а) и размера капель (б) при примерно равном соотношении между количествами сконденсированных олова и алюминия с различной толщиной пленки Al : ○ – 30–60; ● – 70–130; ◕ – 170–210 нм*

Установлено, что на массивном (~200 нм) алюминии угол смачивания θ = 30° для частиц олова размером в несколько микрометров. При толщине пленки алюминия 100 нм угол θ возрастает до 50°. Для еще более тонкой пленки алюминия угол θ воз-растает до 140°, что соответствует углу смачивания оловом окиси алюминия.

Исследовано также влияние на угол смачивания давления остаточных газов в интервале $10^{-3}$–$10^{-9}$ мм рт. ст. в процессе препарирования пленок алюминия и последующей конденсации на них олова. В вакууме, создаваемом при помощи безмасляных средств откачки, при указанных выше условиях получения образцов при толщинах пленок алюминия более 80 нм олово смачивает алюминий вплоть до давления $10^{-4}$ мм рт. ст.; при этом угол $\theta$ с повышением давления монотонно возрастает (измерения проводились для капель олова размером несколько микрометров). В случае конденсации в вакууме $10^{-5}$ мм рт. ст., полученном с помощью масляных диффузионных насосов, при том же режиме препарирования значение краевого угла составляет $\theta = 120$–$140°$. Хорошее смачивание удается получить, если конденсация олова начинается за несколько секунд до прекращения осаждения алюминия. Эти данные наглядно показывают, что даже при конденсации в вакууме $10^{-7}$–$10^{-8}$ мм рт. ст. на смачивание существенное влияние оказывает не только толщина пленки алюминия, но и тончайшая пленка окиси алюминия, образующаяся в этих условиях.

Таким образом, проведенные исследования показывают [15], что в системах с промежуточным слоем изменение краевого угла с толщиной пленки происходит в крайних пределах, соответствующих смачиванию чистой подложки и смачиванию материала пленки в массивном состоянии. Значение критической толщины, ниже которой наблюдается изменение $\theta$, является индивидуальной характеристикой тройной системы и определяется характером взаимодействия компонентов. Для систем с отсутствием химического взаимодействия изменение смачивания происходит при малых толщинах пленок (порядка не-скольких нанометров) и обусловлено размерной зависимостью поверхностной энергии пленок и их несплошностью. Наличие в системе растворения или химического взаимодействия компонентов приводит к сдвигу $t_k$ в область бóльших толщин (десятки нанометров). В этом случае зависимость $\theta(t)$ связана с изменением степени гетерогенности смачиваемой поверхности с уменьшением толщины промежуточного слоя.

В работе [51] выполнен анализ полученных результатов [17, 28, 41–43] совместно с данными [38–40] по смачиванию в тройных системах, который позволил выделить основные типы зависимостей $\theta(t)$ для смачивания расплавом тонкой пленки на поверхности массивной подложки (рис. 28):

а) **невзаимодействующие системы** – рис. 28а. Значение $t_k$ в таких системах определяется микроструктурой промежуточной пленки и в некоторых пределах может зависеть от технологических параметров ее получения (температура подложки, скорость конденсации и др.). Изменение краевого угла определяется переходом от несплошной к сплошной пленке и зависимостью ее поверхностной энергии от толщины. Примерами таких систем могут быть Sn/C/KCl, Sn/Al$_2$O$_3$/KCl (рис. 21), Sn/C/Ge [40];

б) **системы с растворением пленки в жидком металле** – рис. 28б. В этом случае на зависимости присутствует еще одна характерная толщина $t_p$, до которой промежуточная пленка полностью растворяется в расплаве. Значение $t_p$ зависит от растворимости материала пленки в жидком металле при данной температуре. В интервале толщин $t_p < t < t_к$ имеет место частичное растворение пленки в жидком металле, которое приводит к ее несплошности, то есть подложка становится гетерогенной. Поскольку растворимость материала пленки в расплаве для исследованных систем ограничена, то степень гетерогенности подложки зависит от толщины пленки, что и обусловливает наблюдаемую зависимость θ(*t*). Зависимости такого типа наблюдаются в системах Pb/Ni/NaCl, Bi/Fe/KCl, Sn/Al/KCl (рис. 22) и для ряда систем, изученных в работах [38–40]: [Cu, Ag, Pb, Sn]/[Mo, V, Fe]/ [кварц, сапфир, графит];

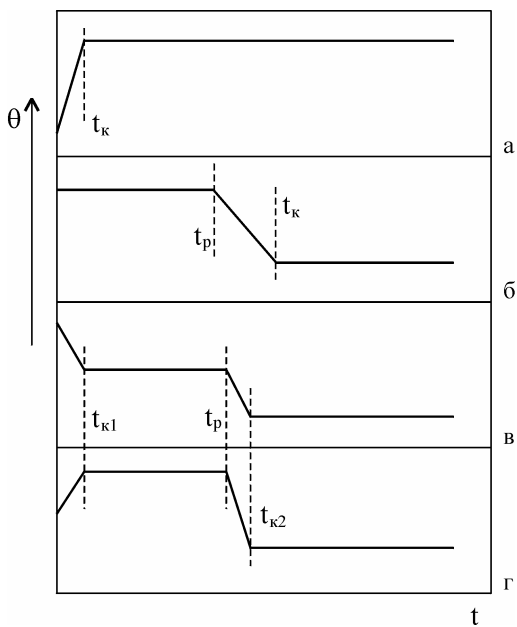

*Рис. 28. Основные типы зависимостей θ(t) для систем расплав – пленка – подложка: а) невзаимодействующие системы; б) системы с растворением пленки в расплаве; в, г) системы с химическим взаимодействием на границе пленка – подложка*

в) **системы с химическим взаимодействием пленки с подложкой** – рис. 28в, г. (Системы Pb/Ni/Si и Pb/Ni/GaAs, рис. 23, 24). Система такого типа может быть разделена на две подсистемы и в соответствии с этим характеризуется двумя значениями критической толщины. Вторая подсистема (области B, C, D на рис. 24) относится к типу а) или б). В первой подсистеме гетерогенность подложки на переходном участке $0 < t < t_{k1}$ (область A на рис. 24) является следствием роста островков новой фазы (химического соединения пленки с подложкой), и величина $t_{k1}$, соответствующая образованию сплошной пленки соединения, определяется механизмом взаимодействия промежуточной пленки с подложкой. Если при взаимодействии происходит образование соединений, которые смачиваются металлом хуже, чем исходные вещества, то возможно появление немонотонных зависимостей $\theta(t)$ (рис. 28г).

Следует отметить, что приведенные на рис. 28 зависимости являются простейшими и влияние других факторов, например, взаимодействия с остаточной атмосферой (Sn/Al/KCl и частично Bi/Fe/KCl), может привести к более сложному изменению краевого угла с толщиной промежуточной пленки.

*Глава 4*

## Смачивание тонких свободных пленок

При интерпретации результатов по смачиванию в трехкомпонентных системах жидкость – тонкая пленка – массивная подложка трудно разделить эффекты, связанные собственно с толщиной пленки и влиянием массивной подложки. Поэтому представлялось [53–55] целесообразным проведение исследований смачивания тонких свободных пленок в зависимости от их толщины. Впервые исследования краевого угла смачивания микрокаплями индия, олова и свинца тонких свободных углеродных пленок с уменьшением их толщины от 30 до 4 нм были выполнены в работах [53, 54]. Полученные результаты, как отмечается в [55], не позволили установить количественную размерную зависимость поверхностной энергии тонких углеродных пленок. Однако эти результаты [55] представляют самостоятельный интерес, поскольку в высокодисперсных системах возможна также ситуация, когда жидкие частицы смачивают поверхность не массивных твердых тел, а свободных тонких пленок. И в этом случае наблюдаются специфические эффекты, связанные с деформацией пленки под жидкой частицей.

В работе [80] была построена теория смачивания упругого полупространства, в соответствии с результатами которой капля деформирует область вблизи линии контакта трех фаз с образованием ранта. В случае тонких пленок деформация может быть значительной, что позволяет обнаружить ее экспериментально и по-лучить из сравнения с теорией важные для практических приложений параметры ультрадисперсных систем. Поэтому ниже кратко рассматривается теоретический анализ смачивания тонких свободных пленок в предположении постоянства поверхностных энергий $\sigma_l$, $\sigma_u$ и $\sigma_{ul}$, выполненный в работе [56], и соответствующие экспериментальные результаты работ [53–55].

## 4.1. Смачивание малыми каплями свободной упругодеформируемой пленки

Согласно [15, 56], равновесные характеристики системы, состоящей из свободной упругодеформируемой пленки толщиной $t$ и смачивающей ее капли (рис. 29), находятся так же, как и в рассмотренной в разделе 2.1 задаче, – из условия минимума свободной энергии, в выражении для которой появляется слагаемое, соответствующее упругой энергии пленки

$$F = 2\pi \int_0^L \left\{ \left[ -p(z - \zeta) + \sigma_l \sqrt{1 - z'^2} + (\sigma_{ul} - \sigma_u)\sqrt{1 + \zeta'^2} \right] \rho \Theta(r - \rho) + \right. \tag{50}$$
$$\left. + 2\sigma_u \rho \sqrt{1 + \zeta'^2} + \psi(\zeta', \zeta'', u, u', \rho) \right\} d\rho,$$

где $L$ – радиус окружности закрепления пленки, функциями $\zeta(\rho)$ и $z(\rho)$ задан радиальный профиль поверхностей пленки и капли соответственно; $\Theta(x)$ – ступенчатая функция Хевисайда. Функция $\psi$ равна сумме вкладов упругих энергий чистого изгиба $\psi_1$ и продольного растяжения пленки $\psi_2$, записываемых, согласно [81], при учете осевой симметрии в следующем виде:

$$\psi_1 = \frac{1}{2} D \rho \left( \zeta'' + \frac{2\nu}{\rho} \zeta'' \zeta' + \frac{1}{\rho^2} \zeta'^2 \right),$$
$$\psi_2 = \frac{6}{t^2} D \rho \left( u'^2 + \frac{2\nu}{\rho} u' u + \frac{u^2}{\rho^2} + \frac{1}{4} \zeta'^4 + u' \zeta'^2 + \frac{\nu}{\rho} u \zeta'^2 \right), \tag{51}$$

где $\nu$ – коэффициент Пуассона, $u$ – радиальная компонента двумерного вектора смещения, а $D = Et^3 / \left[12\left(1 - \nu^2\right)\right]$ – коэффициент жесткости ($E$ – модуль Юнга).

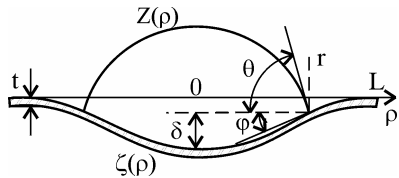

*Рис. 29. Схематическое изображение жидкой капли
на тонкой упругой пленке*

Форма свободной поверхности жидкости находится варьированием функционала $F$ по $\delta z$ в области $0 \le \rho \le r$. Соответствующее уравнение Эйлера после двукратного интегрирования дает функцию $z(\rho)$ в виде сферы (27) радиусом $R = 2\sigma_l / p$.

Варьирование $F$ по $\delta\xi$ и $\delta u$ позволяет получить уравнения, определяющие деформацию пленки, которые после частичного интегрирования и подстановки функций $\psi_1$ и $\psi_2$ из выражения (51) принимают вид

$$\zeta''' + \frac{1}{\rho}\zeta'' - \frac{1}{\rho^2}\zeta' - \frac{12}{t^2}\left(u' + \frac{\nu}{\rho}u + \frac{1}{2}\zeta'^2\right) - $$
$$- \frac{1}{D}\left[(\sigma_{sl} - \sigma_s)\Theta(r - \rho) + 2\sigma_s\right]\frac{\zeta'}{\sqrt{1 + \zeta'^2}} = -\frac{pr}{2D}\Theta(r - \rho); \tag{52}$$

$$u'' + \frac{1}{\rho}u' - \frac{1}{\rho^2}u = -\zeta'\zeta'' - \frac{1 - \nu}{2\rho}\zeta'^2. \tag{53}$$

Граничные условия к уравнениям (52) и (53) вытекают из равенства нулю неинтегральных слагаемых вариации $\delta F$. Два из них были использованы при получении уравнений (52), (53), а остальные могут быть записаны следующим образом: $u(0) = 0;$ $\xi'(0) = 0;$ $\xi(L) = 0;$ $\xi'(L) = 0;$ $u(L) = 0;$ кроме того, в точке $\rho = r$ следует потребовать непрерывность функций $\zeta(\rho)$, $\zeta'(\rho)$, $u(\rho)$, $\zeta''(\rho)$ и $u'(\rho)$.

Условие для равновесного значения краевого угла $\theta$ можно определить путем варьирования функционала (50) по $\delta r$. При этом получается выражение, являющееся уравнением Юнга, записанным вдоль оси, параллельной участку пленки в окрестности точки $\rho = r$:

$$\sigma_l \cos(\theta - \varphi) = \sigma_u - \sigma_{ul}, \tag{54}$$

где $\varphi = \operatorname{arctg}\zeta'(r)$ – угол наклона пленки в точке $\rho = r$. Такое же уравнение Юнга с поправкой на угол наклона упругой поверхности получено и в работе [80].

Еще одно соотношение, связывающее $\varphi$ и $\theta$, следует из уравнения (52) и граничных условий в точке $\rho = r$:

$$\sigma_l \cdot \sin \theta = (\sigma_{ul} - \sigma_u)\sin \varphi + D[\zeta'''(r+0) - \zeta'''(r-0)]. \tag{55}$$

Как видно из (52) и (55), краевой угол зависит от деформации пленки и определяется скачком третьей производной $\zeta(\rho)$ на линии трехфазного контакта.

Была выполнена [15, 56] оценка поведения системы для малых и больших прогибов пленки, т. е. когда преобладающими являются деформации изгиба и растяжения соответственно. В случае, когда максимальный прогиб пленки $\delta$ меньше ее толщины, уравнения (52) и (53) линеаризуются, и из их решения следует соотношение

$$\frac{E}{1 - v^2} = \frac{9\sigma_l \cdot \sin \theta_\infty}{8t^3}\left(\lim_{r \to 0} \frac{\delta}{r^3}\right)^{-1}, \varphi \sim \frac{4}{3}\frac{\delta}{r}, \tag{56}$$

связывающее модуль Юнга с экспериментально измеряемыми параметрами. При больших изгибах пленки ($\delta > t$) могут быть получены приближенные решения уравнений (52) и (53), из которых следует оценка

$$\frac{\delta}{r} \sim \left(2\sigma_l \cdot \sin\theta_\infty / Et\right)^{\frac{1}{3}}. \tag{57}$$

В случае очень тонкой пленки $t \le 10\sigma/E$ (это может иметь место для эластичных пленок с малым модулем упругости) ее форма под каплей стремится к сферической с радиусом $R_{ul} = r(\sigma_u + \sigma_{ul})/\sigma_l \sin\theta$, а вне капли – остается плоской. Гладкий переход от одной формы к другой осуществляется в узком участке шириной порядка толщины пленки, а значение краевого угла $\theta = \lim_{t\to 0}\theta(t)$ определяется только поверхностными энергиями фаз в соответствии с уравнениями

$$\begin{aligned}
\sigma_l \cos\theta_0 + \left(\sigma_u + \sigma_{ul}\right)\cos\varphi &= 2\sigma_u; \\
\sigma_l \sin\theta_0 &= \left(\sigma_u + \sigma_{ul}\right)\sin\varphi.
\end{aligned} \tag{58}$$

## 4.2. Смачивание островковыми конденсатами свободных углеродных пленок

Как уже отмечалось, экспериментальные исследования смачивания островковыми вакуумными конденсатами средней дисперсности свободных аморфных углеродных пленок различной толщины с целью получения сведений о поверхностной энергии пленок были выполнены в работах [53, 54, 55]. Выбор объектов, наряду с методическими соображениями, обусловлен тем, что пленки углерода широко используются в тонкопленочной технологии, а углеродные волокна, пропитанные различными металлами и сплавами, являются основой многих композиционных материалов. Для исследований использовались металлы (In, Sn, Pb), химически инертные по отношению к углероду и образующие с углеродными пленками углы смачивания 140°–150°.

Образцы для исследований препарировались путем испарения и конденсации углерода и металлов в вакууме $10^{-6}$–$10^{-5}$ мм рт. ст., создаваемом безмасляной системой откачки. На сколы монокристаллов KCl в вакууме испарением из вольтовой дуги нано-сились тонкие углеродные пленки. Взаимное расположение испарителя и подложек позволяло получать в одном эксперименте серию образцов с существенно различной толщиной углеродного слоя. Поскольку пленки, полученные при разных углах конденсации, могут иметь различную структуру, подложки располагались перпендикулярно молекулярному пучку углерода. Полученные пленки углерода отделялись от KCl и помещались на медные сеточки с размером ячеек 60 мкм. На приготовленные таким образом свободные углеродные пленки в вакууме $3 \cdot 10^{-6}$ мм рт. ст. путем термического испарения наносились островковые конденсаты олова, свинца и индия. Для обеспечения конденсации исследуемого металла по механизму пар→жидкость температура подложек во время эксперимента поддерживалась выше температуры плавления металла и составляла 520 K для олова, 470 K для индия и 620 K для свинца. Для измерения толщины углеродных пленок использовались метод линий равного хроматического порядка, непосредственное определение толщины на электронно-микроскопических снимках свернувшихся участков пленки и расчет по известной геометрии препарирования образцов. Результаты измерений, полученные этими методами, совпадают, и поэтому, как правило, использовался электронно-микроскопический метод, отличающийся простотой и наглядностью. Погрешность измерения толщин этим методом составляла $\sim 1{,}5$ нм для пленок толщиной 10–20 нм. Толщина пленок углерода изменялась в пределах 4–30 нм, а размер жидких металлических частиц составлял 30–500 нм и ограничивался, с одной стороны, необходимостью исключить размерный эффект, связанный с дисперсностью жидкой фазы, а с другой – прочностью углеродных пленок. Краевые углы смачивания определялись электронно-микроскопически по методу свертки.

При электронно-микроскопических исследованиях профилей закристаллизовавшихся капель металла (рис. 30) было обнаружено существенное различие в форме межфазной границы капля – подложка для микрочастиц, конденсированных на свободных пленках и на пленках, находящихся на твердой поверхности [53, 54, 55].

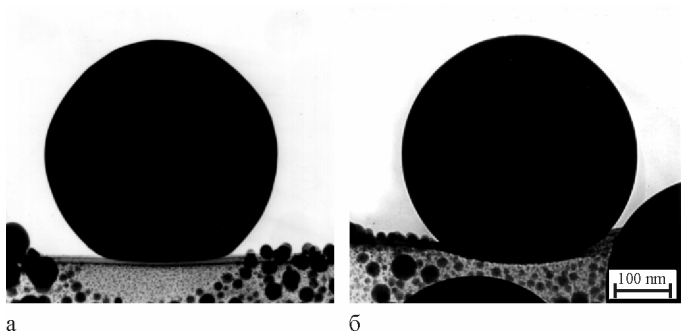

а                              б

*Рис. 30. Микроснимки капель олова на свободных углеродных пленках*
*толщиной 20 (а) и 10 (б) нм [55]*

Это различие заключается в том, что в случае, когда пленка достаточно тонкая, она деформируется каплей (рис. 30), в то время как у частиц, конденсированных на твердой поверхности, граница раздела жидкость-подложка остается плоской. Угол смачивания для капли на тонкой пленке определялся как угол между плоскостью пленки и касательной к жидкой поверхности в точке тройного контакта.

Параметры, характеризующие форму частиц (радиус кривизны поверхности капли $R$, радиус основания $r$ и ее высота $H$), измерялись на микроснимках свернувшихся участков пленки с закристаллизовавшимися частицами. Согласно выполненным исследованиям микрокапли имеют форму сферического сегмента, и поэтому для нахождения $\theta$ использовались соотношения   $\theta_1 = 2\mathrm{arctg}(H/r);$   $\theta_2 = \arccos(1 - H/R)$   и $\theta_3 = 180° - \arcsin(H/r)$, при $\theta > 90°$. Для сферических частиц величины $\theta_1$, $\theta_2$ и $\theta_3$ должны быть равны, однако, так как при кристаллизации возможно некоторое искажение формы капель (вследствие появления огранки, а также изменения объема при затвердевании), эти величины несколько различны, и поэтому для повышения точности определения углов брались средние значения $\overline{\theta} = (\theta_1 + \theta_2 + \theta_3)/3$. Погрешность определения краевого угла находилась из соотношения $\Delta\theta = \Delta\theta_i + \Delta\theta_s$, где $\Delta\theta_i = (1-2)°$ – инструментальная ошибка, а $\Delta\theta_s = \sqrt{(\theta_1 - \overline{\theta})^2 + (\theta_2 - \overline{\theta})^2 + (\theta_3 - \overline{\theta})^2}$ – погрешность, обусловленная отклонением формы закристаллизовавшейся капли от сферической. При фиксированной толщине углеродной пленки измерялись краевые углы для 20–30 частиц, и среднее значение $\theta$ для данного образца определялось с учетом погрешностей [55] как $\theta = \sum\limits_{i=1}^{n} \dfrac{\theta_i}{\Delta\theta_i} \Big/ \sum\limits_{i=1}^{n} \dfrac{1}{\Delta\theta_i}$, где $n$ – количество капель ($i = 1, 2, ..., n$).

Для исследованных систем было установлено [55], что при $t < 30$ нм краевой угол уменьшается с толщиной пленки (рис. 31). При $t > 30$ нм угол $\theta$ приближается к постоянному значению $\theta_\infty$, соответствующему смачиванию массивного материала. Анализ профилей микрокапель на свободных пленках показал, что пленка деформируется каплей, при этом величина прогиба зависит от толщины и становится существенной при $t < 10$ нм.

Полученные результаты были интерпретированы [15, 56] в рамках теории смачивания упругодеформируемых пленок [56], основные положения которой изложены в разделе 3.1. Как следует из соотношений (56) и (57), при $t = \text{const}$ зависимость максимального прогиба $\delta$ пленки от радиуса основания капли $r$ будет различной в случаях, когда преобладающими являются деформации изгиба ($\delta \sim r^3$ при $\delta < t$) и растяжения ($\delta \sim r$ при $\delta > t$).

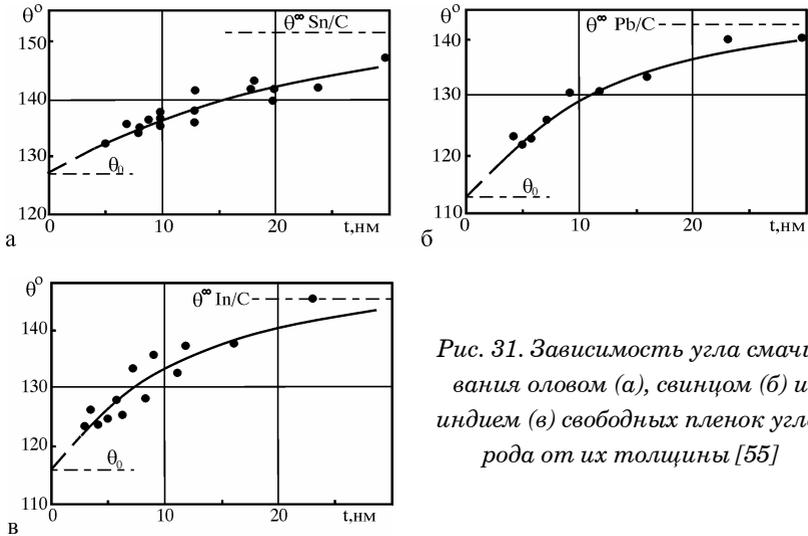

а

б

в

*Рис. 31. Зависимость угла смачивания оловом (а), свинцом (б) и индием (в) свободных пленок углерода от их толщины [55]*

Экспериментальные зависимости $\delta(r)$ для системы Sn/C при $t = 10$ нм (рис. 32) подтверждают этот вывод. Из приведенных графиков видно, что линейная зависимость $\delta(r)$ наблюдается при $\delta < t$ в координатах «$r^3 - \delta$», а при $\delta > t$ – в координатах «$r - \delta$».

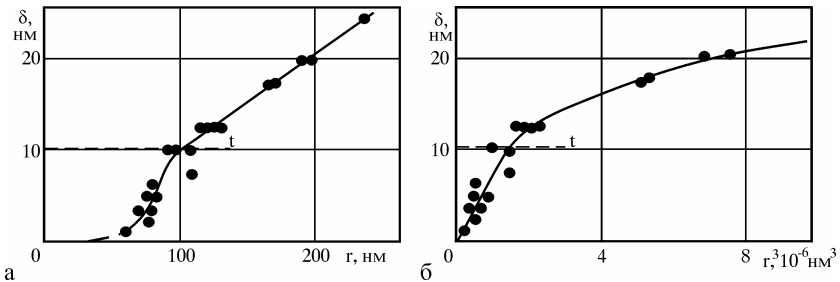

а

б

*Рис. 32. Зависимость $\delta(r)$ в координатах "$r - \delta$" (а) и "$r^3 - \delta$"(б) для капель олова на свободной пленке углерода толщиной 10 нм [15, 55].*

Из этих же графиков следует, что $\left.\left(\delta/r\right)_{\delta>t}\right|_{\text{exp.}} = 0{,}11$ и $\left[\left(\lim\limits_{r\to 0}\delta\big/r^3\right)_{\delta<t}\right]_{\text{exp.}} = 7{,}5\cdot10^{-6}$ нм$^{-2}$. Поскольку коэффициент Пуассона обычно находится в пределах $1/4 < \nu < 1/2$, то из соотношения (56) можно оценить модуль Юнга углеродной пленки, который оказывается равным $E \approx (3{,}4\text{--}4{,}2)\cdot10^{10}$ Н/м$^2$. С учетом этого значения из (57) получается $\left.\left(\delta/r\right)_{\delta>t}\right|_{\text{theor.}} \approx 0{,}10 \pm 0{,}01$, что хорошо согласуется с экспериментом. Теоретическая оценка краевого угла при $t = 10$ нм также дает значение $\theta_{\text{theor.}} = (140 \pm 1)°$, близкое к экспериментальному $\theta_{\text{exp.}} \approx 138°$.

Результаты теоретического рассмотрения для очень тонких пленок были использованы для оценки поверхностной энергии углеродных пленок [15]. Как следует из (58) и (18), в случае, когда энергией деформации по сравнению с поверхностной энергией можно пренебречь, значение $\sigma_u$ определяется выражением

$$\sigma_u = \sigma_l \sin\theta_\infty \big/ 4\big(\cos\theta_0 - \cos\theta_\infty\big). \qquad (59)$$

Интервал толщин свободных пленок, в котором выполняется это соотношение, является экспериментально недостижимым ($t < 1$ нм), однако угол $\theta_0$ может быть найден экстраполяцией зависимости $\theta(t)$. Величины поверхностной энергии углеродной пленки, оцененные с помощью (59) из экспериментальных данных для систем In/C, Sn/C и Pb/C (табл. 2), составляют примерно $120 \pm 30$ мДж/м$^2$ и согласуются как между собой, так и с имеющимися в литературе значениями.

Поскольку сведения о поверхностной энергии аморфных углеродных пленок отсутствуют, в работе [55] полученные значения $\sigma_u$ сравниваются с имеющимися данными о поверхностной энергии различных модификаций углерода. Величины $\sigma_u$ углерода, найденные различными авторами, оказываются в весьма широком интервале $34{,}6\text{--}2560$ мДж/м$^2$.

**Результаты по смачиванию свободных тонких пленок углерода [55]**

| Система | $\theta_\infty, °$ | $\theta_0, °$ | $\sigma_l$, мДж/м$^2$ | $\sigma_u$, мДж/м$^2$ |
|---------|------------|---------|---------|---------|
| Pb/C | 142 | 113 | 450 | $113 \pm 30$ |
| In/C | 145 | 116 | 559 | $121 \pm 30$ |
| Sn/C | 151 | 127 | 540 | $118 \pm 30$ |

Ряд значений $\sigma_u$, рассчитанных ($\sigma_u \approx 170$ мДж/м$^2$ [57, 58]) или определенных экспериментально с помощью различных методов ($\sigma_u = 110,6$ мДж/м$^2$ [59] – смачивание; 110–120 мДж/м$^2$ [2] – смачивание, адсорбция, теплота растворения; 119 мДж/м$^2$ [60] – теплота смачивания; 130 мДж/м$^2$ [61] – уширение дислокационных рядов; 110–160 мДж/м$^2$ [62] – кинетика графитизации сплавов), неплохо согласуются с результатами работы [55]. Величины $\sigma_u = 46$ мДж/м$^2$ [49] и $\sigma_u = 34,6$ мДж/м$^2$ [64], найденные из данных по смачиванию водой и органическими жидкостями различных типов графита и застеклованного углерода, вероятно, занижены вследствие адсорбции молекул жидкости на твердой поверхности [3].

Большие значения поверхностной энергии поликристаллического графита, полученные в работах [65, 66] методом «нейтральной капли», объясняются следующим образом. Вследствие сильной анизотропии кристаллической решетки графита, значение $\sigma_u$ для базисной ($\sigma^b$) плоскости и боковой грани призмы ($\sigma^s$) существенно различны. Отношение $\sigma^b/\sigma^s$, найденное в работе [67] из теоремы Вульфа, оказывается весьма значительным и составляет $\sigma^b/\sigma^s = 13$. Следовательно, величины $\sigma_u = 2500$ мДж/м$^2$ [65] и $\sigma_u = 2580$ мДж/м$^2$ [66] обусловлены наличием на поверхности образца поликристаллического графита зерен с различной ориентацией. В связи с этим отмечается также обнаруженное в работе [68] изменение поверхностной энергии пирографита от 2000 до 350 мДж/м² в зависимости от степени совершенства структуры образца и его предыстории (температуры и длительности отжига). Кроме того, величина $\sigma_u = 350$ мДж/м² [68], равно как и расчетное значение $\sigma_u = 506$ мДж/м² [69], вероятно, не могут быть приняты в качестве характеристик аморфной углеродной пленки, так как вычисление $\theta_0$ при $\sigma_u = 350$ мДж/м² приводит к величинам (~ 130, 140 и 135° соответственно для In, Sn и Pb), превышающим измеренные экспериментально.

Анализируя данные по смачиванию островковыми конденсатами индия, олова и свинца свободных углеродных пленок, необходимо отметить следующее. Как показывают количественные оценки этих результатов, изменение $\theta$ с толщиной свободных пленок вследствие предсказываемой теоретически размерной зависимости их поверхностной энергии примерно на порядок меньше изменения краевого угла из-за деформации. Поэтому проведенные исследования не позволили проследить зависимость $\theta(t)$ для свободных углеродных пленок, но они дали возможность определить значение поверхностной энергии для них, что невозможно сделать другими методами.

*Глава 5*

## Смачивание в переохлажденных островковых конденсатах

Известно, что выше температуры плавления поверхностная энергия $\sigma_l$ линейно уменьшается с повышением температуры. Однако существующие представления и экспериментальные данные о температурной зависимости поверхностной энергии жидкостей неоднозначны, о чем свидетельствуют приведенные в работе [70] результаты и их анализ. При этом в работе [70] высказывается предположение, что при значительных величинах переохлаждений можно ожидать инверсию температурной зависимости $\sigma_l(T)$ (т. е. переход к $d\sigma_l/dT>0$), обусловленную различным изменением с температурой поверхностной и объемной энтропии.

Экспериментально температурная зависимость поверхностной энергии переохлажденных металлов (Ga, In, Sn, Bi, Pb) [71] изучена лишь в области небольших переохлаждений. При этом установлено, что зависимость $\sigma_l(T)$ линейна ($d\sigma_l/dT<0$) и сохраняется при переходе в область переохлажденного состояния. В то же время в работе [71] указывается на незначительное отклонение от линейности $\sigma_l(T)$ ниже температуры плавления олова и висмута, для которых получены переохлаждения 42 и 59 K соответственно. Это, однако, оставляет открытым вопрос о существовании инверсии температурной зависимости поверхностной энергии переохлажденной жидкости, так как авторам [71] не удалось достичь больших переохлаждений. В то же время согласно [70] не исключено, что при достаточном переохлаждении у всех жидких металлов будет наблюдаться инверсия.

Измерения поверхностной энергии переохлажденных расплавов сложны, так как значительные переохлаждения достигаются обычно в микрообъемах, а при традиционных методах определения $\sigma_l$ требуется большое количество расплава [1, 3]. Сведения о зависимости поверхностной энергии от температуры при $T < T_s$ можно получить, исследуя смачивание переохлажденным расплавом твердой подложки с применением вакуумных конденсатов. Это связано с тем, что для последних достаточно просто получить большие переохлаждения и, используя соответствующие методы [12], определить краевые углы смачивания с достаточной точностью.

Такие эксперименты впервые были выполнены в работах [72–76], в которых исследовались контактные пары, представляющие собой островковые пленки олова, индия, висмута и меди на аморфных углеродных подложках и индия на алюминиевой подложке. Выбор систем обусловлен тем, что указанные металлы не образуют химических соединений с используемой подложкой и практически не растворяют ее. Кроме того, при конденсации в вакууме олова и индия на углеродных подложках достигаются большие переохлаждения ($\Delta T \geq T_s/3$). В то же время в работе [70] отмечается, что у олова инверсия отсутствует, а для индия, если бы его можно было достаточно сильно переохладить, можно было бы наблюдать инверсию.

### 5.1. Инверсия температурной зависимости смачивания в островковых пленках

Для изучения температурной зависимости смачивания использовалась специально разработанная методика [73, 75], позволяющая в одном эксперименте определять зависимость $\theta(T)$ в широком температурном интервале: от температуры максимального переохлаждения на выбранной подложке $T_g$ до критической температуры конденсации $T_k$. Образцы для исследований препарировались конденсацией в вакууме $5 \cdot 10^{-6}$–$2 \cdot 10^{-8}$ мм рт. ст. на круговую подложку, вдоль которой устанавливался градиент температур (200–900 K). В качестве подложки применялась пластина из нержавеющей стали с укрепленными на ней монокристаллами NaCl или полированными пластинками $Al_2O_3$, на которые непосредственно перед конденсацией исследуемого металла наносилась углеродная пленка толщиной 20–30 нм. В результате на подложке в интервале температур $T_g < T < T_k$ в соответствии с диаграммой конденсации [77, 78] происходит конденсация в равновесную или переохлажденную жидкую фазу с образованием островков (микрокапель), являющихся подобными сферическими сегментами. Полученные образцы охлаждались в вакууме до комнатной температуры, затем измерялись краевые углы смачивания на закристаллизовавшихся каплях, сконденсированных при различных температурах подложки. Вследствие гистерезиса смачивания [33], который даже на абсолютно гладкой и однородной поверхности возникает благодаря деформации подложки в зоне тройного контакта [34], при изменении температуры радиус основания капель оказывается постоянным. Это подтверждается исследованиями, выполненными непосредственно в электронном микроскопе: при нагреве капель вплоть до температур, при которых начинается их испарение, и при их охлаждении до комнатной температуры не наблюдается срыва периметра смачивания. Поскольку радиус основания капли остается неизменным и капля сохраняет сферическую форму, изменение наблюдаемого значения краевого угла при охлаждении возможно только вследствие изменения ее объема. Оценка возможного изменения $\theta$ из-за темпера-турного сжатия металла и скачка объема при кристаллизации дает значение $\leq 2°$ при охлаждении от 700 К до комнатной температуры [12], что меньше погрешности измерений $\pm 3°$. Поэтому величины θ, измеренные на закристаллизовавшихся каплях, правомерно относить к температурам их образования в процессе конденсации. Краевые углы измерялись методом свертки на электронно-микроскопических снимках профилей частиц (рис. 33) и усреднялись для 10–20 частиц, конденсированных при фиксированной температуре. Поскольку конденсация осуществляется на подложку с градиентом температуры, то зависимость θ($T$) может быть измерена в одном эксперименте, т. е. в идентичных условиях получения капель в широком интервале температур и со сколь угодно малым температурным шагом.

Результаты измерений краевых углов смачивания в системах Sn/C и In/C [73] приведены на рис. 34. Полученные зависимости характеризуются максимумом при температурах 550 и 500 К для олова и индия соответственно. Ниже $T_s$ угол смачивания плавно уменьшается с понижением температуры. Уменьшение θ для исследованных систем составляет примерно 25° при максимальных достигнутых переохлаждениях $\Delta T_{\mathrm{Sn}} = 160$ К и $\Delta T_{\mathrm{In}} = 100$ К. Зависимость θ($T$) для индия несколько смещена относительно кривой θ($T$) для олова в область более низких температур, что представляется естественным следствием различных температур плавления этих металлов. Улучшение смачивания наблюдается также и выше $T_s$ при возрастании температуры, причем для индия и олова θ уменьшается в одном и том же температурном интервале $550 < T < 650$ К. Выше 700 К краевой угол в системе Sn/C обнаруживает поведение, типичное для невзаимодействующих систем, заключающееся в слабом уменьшении θ с ростом температуры [1].

При этом зависимость θ(*T*) близка к линейной с коэффициентом наклона $d(\cos\theta)/dT \approx 0{,}0001$ К$^{-1}$.

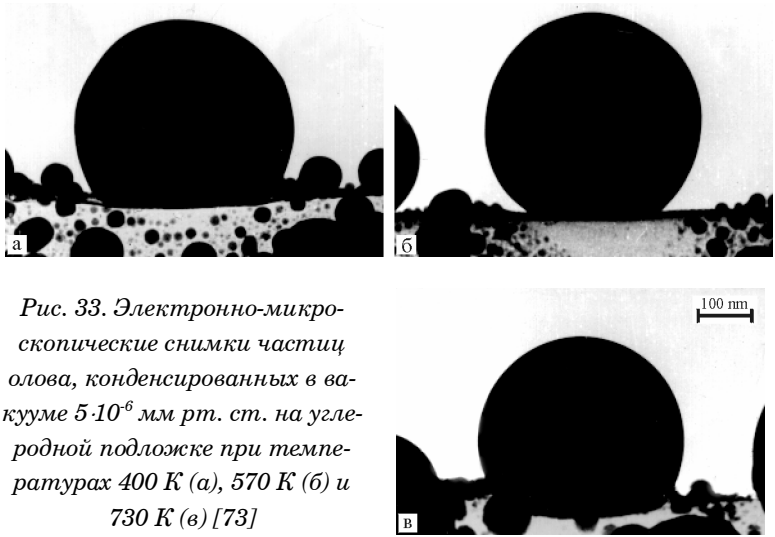

*Рис. 33. Электронно-микроскопические снимки частиц олова, конденсированных в вакууме $5 \cdot 10^{-6}$ мм рт. ст. на углеродной подложке при температурах 400 К (а), 570 К (б) и 730 К (в) [73]*

В системе Bi/C [74] температурная зависимость смачивания, как и в рассмотренных выше системах In/C и Sn/C [73, 75], является немонотонной и характеризуется значительным понижением краевого угла при приближении к температуре максимального переохлаждения (рис. 35а). Однако максимальное значение θ для висмута достигается при $T = 430$ К, то есть в переохлажденном состоянии, в отличие от олова и индия, для которых максимум зависимости θ(*T*) находится выше температуры плавления. Этот факт, а также то, что в данных экспериментах для висмута получены меньшие относительные переохлаждения ($\Delta T/T_s \approx 0.27$), чем для олова ($\Delta T/T_s \approx 0.38$), приводят к тому, что интервал уменьшения краевого угла смачивания с понижением температуры оказывается достаточно узким: $400 < T < 420$ К. В этом интервале происходит уменьшение θ на 25°, что отвечает уменьшению адгезионного натяжения на 50%, т. е. от 280 до 140 мДж/м$^2$ (для поверх-ностной энергии висмута принято значение $\sigma_l = 385$ мДж/м$^2$, полученное экстраполяцией данных [71] до температуры 400 К).

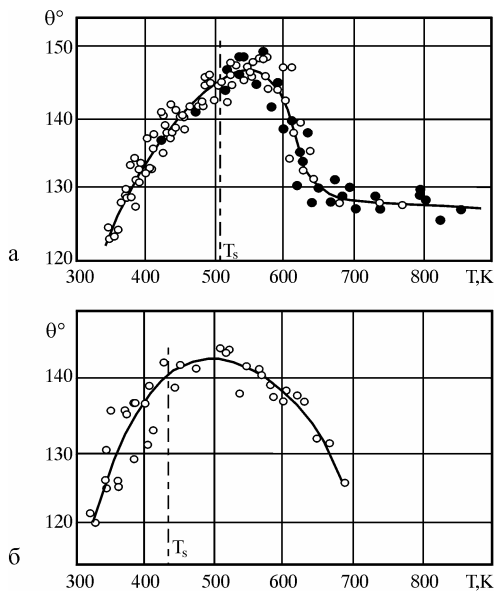

*Рис. 34. Температурные зависимости угла смачивания островковыми*
*конденсатами олова (а) и индия (б) углеродных пленок*
*(подложка-основание: ○ – NaCl; ● – Al$_2$O$_3$; вакуум 5·10$^{-6}$ мм рт. ст.) [73]*

Система In/Al также характеризуется немонотонной зависимостью краевого угла от температуры (рис. 35б), которая подобна зависимости $\theta(T)$ в системе Sn/C [73, 75], однако почти полностью находится выше температуры плавления индия. Для In/Al получены небольшие относительные переохлаждения ($\Delta T/T_s \approx 0{,}05$), что вообще характерно для конденсатов металлов на металлических подложках [77, 78]. Начиная с температуры $T = 420$ К ($\theta = 60°$), краевой угол увеличивается с ростом температуры и при $T = 490$ К принимает максимальное значение $\theta = 143°$. Далее следует быстрое уменьшение $\theta$, и при $T > 500$ К краевой угол имеет постоянное значение $\theta = 120°$. Характерно, что при температуре плавления и ниже ее, в переохлажденном состоя-нии, индий смачивает алюминиевую подложку. Переход от смачивания к несмачиванию, т. е. изменение знака адгезионного натяжения, в системе In/Al наблюдается при $T = 440$ К.

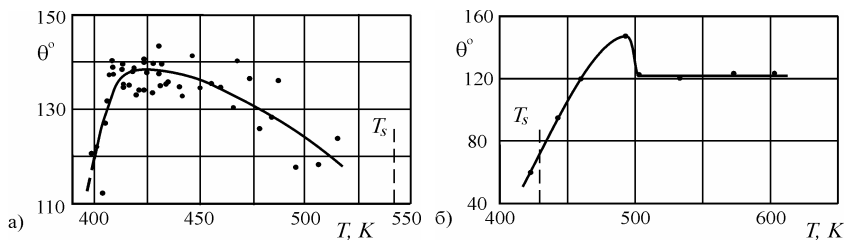

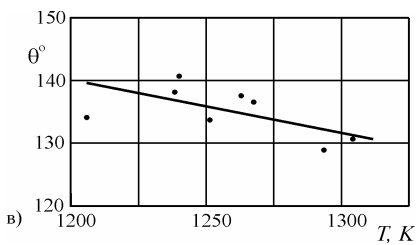

*Рис. 35. Температурные зависимости смачивания для переохлажденных островковых конденсатов металлов на различных подложках. Системы Bi/C (а), In/Al (б), Cu/C (в) [74]*

Для системы Cu/C (рис. 35в) температурная зависимость смачивания не имеет каких-либо особенностей: в интервале температур $1200 < T < 1300$ К наблюдается уменьшение краевого угла с ростом температуры ($d(\cos\theta)/dT \approx 0{,}001$ К$^{-1}$). Это, с одной стороны, подобно поведению $\theta(T)$ для системы Bi/C при тех же величинах относительных переохлаждений, а с другой – линейная зависимость $\theta(T)$ является типичной для контактных систем с невзаимодействующими компонентами [1, 3].

Наблюдаемые изменения краевого угла в области переохлажденного состояния металла, как отмечается в работе [76], вероятно, обусловлены аномальным поведением либо поверхностной энергии жидкого металла, либо межфазной энергии границы металл – углерод. Если предположить $\sigma_{sl}$ неизменной или возрастающей с понижением температуры, то в соответствии с уравнением Юнга экспериментальные данные по $\theta(T)$ свидетельствуют о резком увеличении поверхностной энергии жидкого металла. Так, для олова при $T \leq 400$ К величина $\sigma_l$, найденная в предположении постоянства адгезионного натяжения, превышает соответствующее значение для твердого металла. Следовательно, кристаллизация олова при $T < 400$ К будет сопровождаться уменьшением поверхностной энергии, что не согласуется с существующими теоретическими представлениями и экспериментальными данными. Таким образом, предположение о постоянстве и, тем более, о возрастании $\sigma_{ul}$ с увеличением переохлаждения приводит к противоречию. Поэтому в работе [76] считается, что из данных по смачиванию в переохлажденном состоянии наиболее вероятно следует значительное уменьшение межфазной энергии границы переохлажденная капля – подложка при понижении температуры. Зависимости $\sigma_{ul}(T)$, рассчитанные с использованием линейной экстраполяции в область больших переохлаждений данных по температурной зависимости поверхностной энергии олова и индия [71], приведены на рис. 36.

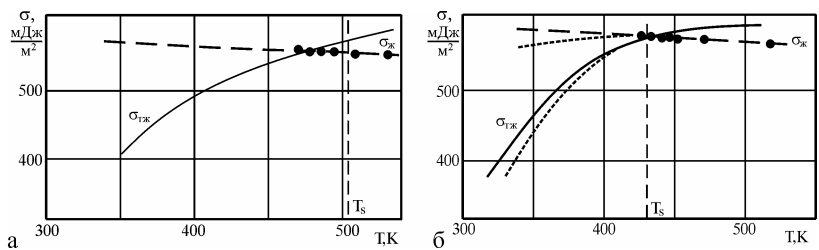

*Рис. 36. Зависимости поверхностной энергии жидкой фазы [71] и межфазной энергии границы капля – подложка от температуры для систем Sn/C (а) и In/C (б). Пунктирные линии на графике (б) соответствуют возможному изменению $\sigma_l$ и $\sigma_{ul}$ с температурой за счет адсорбции остаточных газов*

Среди причин, вызывающих столь значительное уменьшение межфазной энергии, в работе [76] указывается на адсорбцию газовых примесей, величина которой возрастает с понижением температуры, инверсию поверхностной энергии металла в переохлажденном состоянии и на уменьшение различия между твердой и жидкой фазами при увеличении переохлаждения. Однако, учитывая то обстоятельство, что для ряда исследованных металлов (In, Sn, Bi) инверсия температурной зависимости смачивания наблюдается примерно в одном и том же температурном интервале, а в системе Cu/C, при высоких температурах, не обнаружена вовсе, вероятно, определяющей следует признать [76] возрастающую при таких температурах адсорбцию примесей из остаточных газов, приводящую к уменьшению поверхностной энергии металла и межфазной энергии на границе с углеродной подложкой при увеличении переохлаждения (пунктир на рис. 36б), а следовательно, и к экспериментально наблюдаемому улучшению смачивания. С повышением температуры подложки выше 500–600 K происходит увеличение $\sigma_u$ углеродной пленки вследствие резкого уменьшения адсорбции газов на ее поверхности, что улучшает смачивание.

## 5.2. Влияние давления и состава остаточной атмосферы на смачивание оловом углеродных подложек

Наблюдаемые экспериментально зависимости $\theta(T)$ (рис. 34, 35) не могут быть объяснены линейным изменением поверхностных энергий контактирующих фаз [1]. Кроме того, одним из возможных объяснений немонотонного хода $\theta(T)$ является влияние адсорбированных газовых примесей. Поэтому в работах [76, 79] были исследованы температурные зависимости краевых углов смачивания для островковых пленок олова на углеродных подложках, конденсированных при контролируемом составе остаточной атмосферы в условиях, когда существенно уменьшается влияние адсорбированных примесей на поверхностную энергию как углеродных пленок, так и капель металла и границы их раздела.

Препарирование образцов осуществлялось с применением изложенной выше методики [73–76] в прогреваемой вакуумной установке с металлическими уплотнениями при давлении остаточных газов $10^{-7}$–$10^{-9}$ мм рт. ст. Для контроля остаточной атмосферы использовался радиочастотный масс-спектрометр, а ее состав изменялся путем напуска требуемого газа в установку, откачанную до давления $10^{-9}$ мм рт. ст., или путем изменения режима работы основного откачивающего насоса (гетероионный насос типа «орбитрон»). Краевые углы смачивания измерялись по электронно-микроскопическим снимкам профилей частиц на свернувшихся (рис. 37) или наклонных (рис. 38) участках углеродной пленки (методы свертки и наклонного наблюдения [12]); значение θ для фиксированной температуры находилось усреднением с учетом погрешностей значений краевых углов для 10–20 микрочастиц.

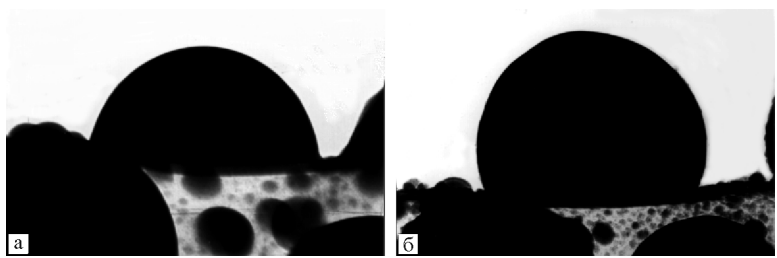

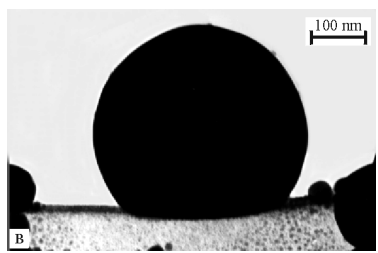

*Рис. 37. Электронно-микроскопические снимки закристаллизовавшихся капель олова, конденсированных в вакууме $2 \cdot 10^{-8}$ мм рт. ст. на углеродных подложках при температурах 350 К (а), 410 К (б) и 500 К (в) [76]*

Результаты измерений θ($T$) для капель олова, полученных в различном вакууме, приведены на рис. 39. Сравнение экспериментальных данных по смачиванию в пленках олова, препариро-ванных в вакууме $10^{-6}$ и $10^{-8}$ мм рт. ст., показывает, что их отличие заключается, во-первых, в отсутствии максимума на зависимости $\theta(T)$ (давление $10^{-8}$ мм рт. ст.) и, во-вторых, указанная зависимость при улучшении вакуума смещается в область меньших значений углов смачивания в температурном интервале переохлажденного состояния капель олова 500–350 K (для давления $10^{-8}$ мм рт. ст. это смещение составляет 20–30°).

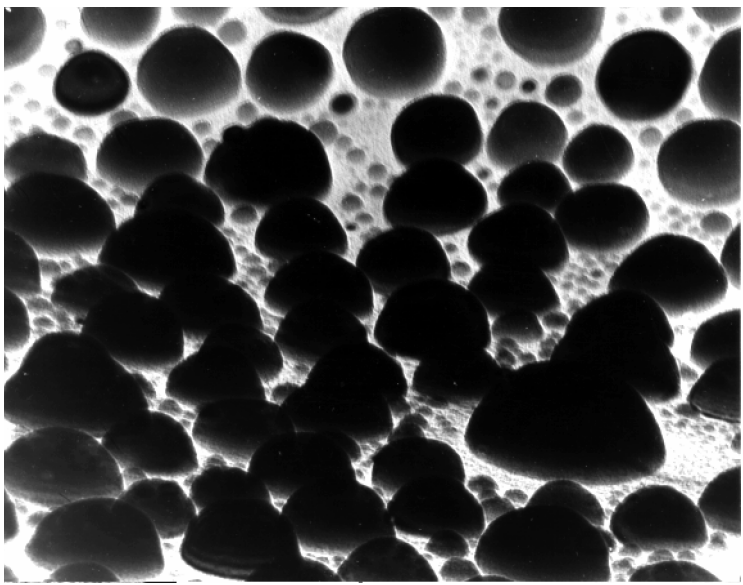

*Рис. 38. Электронно-микроскопический снимок островковой пленки олова, конденсированной в вакууме $2 \cdot 10^{-8}$ мм рт. ст. на углеродной подложке при температуре 315 K  ($\theta=82°$) [79]*

При температурах подложки $T > 500$ K для пленок, препарированных в вакууме $10^{-8}$ мм рт. ст., зависимость $\theta(T)$ выходит на постоянное значение, и углы смачивания оказываются приблизительно равными углам $\theta$ для пленок, полученных при $p = 10^{-5}$ мм рт. ст., но при температурах выше 650 K. Это указывает на то, что максимум на температурной зависимости смачивания для островковых конденсатов олова, индия и вис-мута, полученных в вакууме $10^{-5}$–$10^{-6}$ мм рт. ст. [73, 75], обусловлен влиянием адсорбированных газовых примесей из остаточной атмосферы, которые соответствующим образом изменяют поверхностные энергии контактирующих фаз. При конденсации же в вакууме $10^{-6}$ мм рт. ст. при температурах выше 650 К происходит десорбция газовых примесей с поверхности подложки, а для пленок, полученных в вакууме $10^{-8}$ мм рт. ст., эти примеси отсутствуют и при более низких температурах, и в результате значения углов смачивания в обоих случаях практически совпадают. Характерно, что зависимость $\theta(T)$ для пленок, конденсированных в вакууме $10^{-7}$ мм рт. ст., занимает промежуточное положение между данными для образцов, полученных при $p = 10^{-6}$ и $10^{-8}$ мм рт. ст.

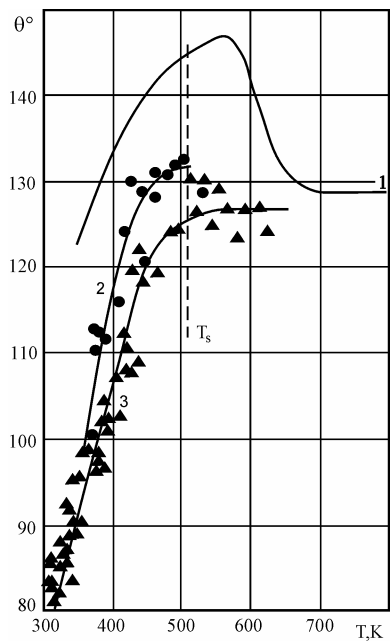

*Рис. 39. Температурная зависимость краевого угла смачивания углеродных подложек островковыми конденсатами олова, препарированными при различных давлениях остаточных газов: 1) 5·10⁻⁶ мм рт. ст. [73], 2) 3·10⁻⁷ мм рт. ст. и 3) 2·10⁻⁸ мм рт. ст. [76]*

Необходимо отметить, что для переохлажденного состояния олова снижение температуры во всех случаях приводит к уменьшению краевых углов. При давлении остаточных газов $10^{-8}$ мм рт. ст. улучшение смачивания оказывается весьма значительным и составляет $\Delta\theta \sim 50°$. Изменение смачивания с температурой хорошо иллюстрируется электронно-микроскопическими снимками профилей частиц (рис. 37) и наклонных участков пленки вблизи температуры максимального переохлаждения (рис. 38). При этом, как можно видеть из графика (рис. 39, кривая 3) и микроснимков (рис. 37, 38), в области глубоких переохлаждений (при $T < 350$ К) наблюдается переход от несмачивания ($\theta > 90°$, межфазная энергия границы капля – подложка $\sigma_{ul}$ превышает поверхностную энергию подложки $\sigma_u$: $\sigma_{ul} > \sigma_u$) к смачиванию ($\theta < 90°$, $\sigma_{ul} < \sigma_u$). Наличие такого перехода, а по существу – изменения знака адгезионного натяжения $\sigma_u - \sigma_{ul}$, подтверждает вывод о значительном уменьшении с температурой межфазной энергии границы переохлажденная капля – подложка, определяющим фактором которого, как отмечается в работе [76], является возрастающая при понижении температуры адсорбция примесей из остаточных газов.

В работе [79] предпринята попытка выяснить, какие именно компоненты остаточной атмосферы ответственны за немонотонное поведение краевого угла. Зависимости (рис. 39) получены в условиях, когда состав атмосферы при конденсации пленок определяется парциальными скоростями откачки различных газов гетеро-ионным титановым насосом типа «орбитрон» с охлаждаемой водой рабочей поверхностью. В соответствии с результатами масс-спектрометрических исследований, основными компонентами такой остаточной атмосферы являются газы метановой группы, азот, кислород, оксид углерода, аргон и вода (для установки без предварительного прогрева). При охлаждении рабочей поверхно-сти жидким азотом изменяются скорости откачки различных газов и основными компонентами атмосферы являются газы метановой группы, которые образуются из двуокиси углерода и воды в процессе работы гетеро-ионного насоса. Температурная зависимость смачивания, соответствующая таким условиям откачки, приведена на рис. 40 (кривая 4).

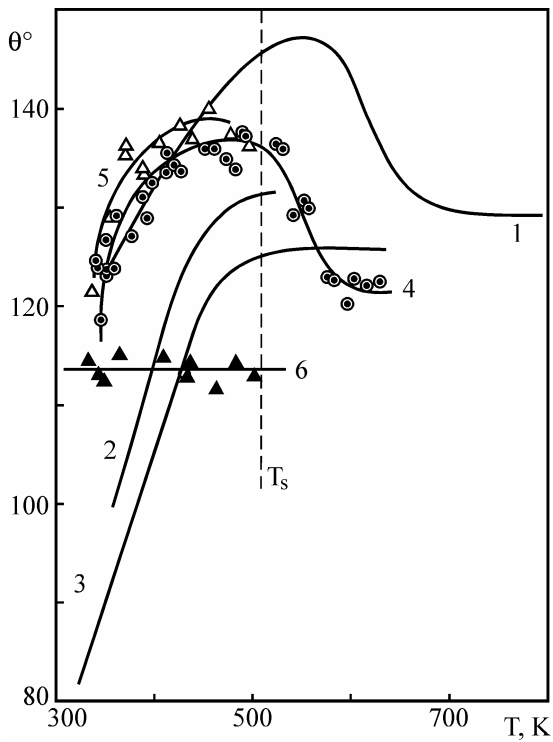

*Рис. 40. Температурные зависимости угла смачивания для частиц олова, конденсированных на углеродной подложке при различном давлении и составе остаточных газов. 1) $10^{-6}$мм рт. ст.; 2) $10^{-7}$ мм рт. ст.; 3) $10^{-8}$ мм рт. ст.; 4) $10^{-9}$ мм рт. ст. (охлаждение насоса жидким азотом); 5) $10^{-5}$–$10^{-6}$ мм рт. ст. (напуск метана); 6) $10^{-5}$–$10^{-6}$ мм рт. ст. (напуск кислорода) [79]*

Полученная зависимость оказалась подобной зависимости $\theta(T)$, соответствующей вакууму $10^{-6}$ мм рт. ст. при охлаждении насоса водой (кривая 1) [73–75], то есть имеет максимум вблизи температуры плавления олова. На этом основании в работе [79] делается вывод, что переход от монотонной зависимости к зависимости с максимумом обусловливается влиянием газов метановой группы. Это предположение также подтверждается [79] измерениями температурной зависимости смачивания для пленок, препарированных при давлении $10^{-6}$ мм рт. ст. в обогащенной метаном атмосфере, которая создавалась путем напуска метана в вакуумную камеру, предварительно откачанную до давления $10^{-9}$ мм рт. ст. Полученные при этом результаты (кривая 5 на рис. 40) хорошо согласуются с данными, отвечающими охлаждению рабочей поверхности насоса жидким азотом при давлении $10^{-9}$ мм рт. ст.

Одним из наиболее активных газов, присутствующих в остаточной атмосфере при конденсации пленок исследуемой системы, является кислород. Поэтому в работе [79] была исследована зависимость $\theta(T)$ при препарировании пленок в атмосфере кислорода при давлении $10^{-6}$ мм рт. ст. Полученные результаты приведены на рис. 40 (зависимость 6). В атмосфере кислорода в исследованном интервале температур $340 < T < 500$ K краевой угол смачивания оказывается приблизительно постоянным, то есть эффект улучшения смачивания с увеличением переохлаждения не наблюдается вообще. Это объясняется [79] тем, что в соответствии с электронографическими данными при конденсации олова в атмосфере кислорода на начальных стадиях образуется оксид олова SnO и полученное значение $\theta \sim 114°$ отвечает смачиванию оловом собственного оксида. При этом, вероятно, не успевают проявляться явления адсорбции и десорбции других газовых примесей и краевой угол практически не зависит от температуры. В то же время, как показывает сравнение имеющихся данных, соответствующих препарированию образ-цов в различных условиях (кривые 1–5 на рис. 40), при конденсации в равновесной атмосфере кислород не оказывает никакого влияния на ход зависимости $\theta(T)$. Это связано с тем, что кислород как химически активный газ эффективно откачивается насосами, работающими на принципе гетерного поглощения, и вследствие этого его концентрация в остаточной атмосфере недостаточна для образования заметных количеств оксидов, а следовательно, и для изменения температурной зависимости смачивания.

Анализ результатов [72–76, 79] позволяет предположить, что собственно переохлажденное состояние металла не является основной причиной резкого улучшения смачивания с понижением температуры для легкоплавких металлов. На это указывает также то, что инверсия температурной зависимости смачивания наблюдается как вблизи температуры плавления (Sn/C, In/C), так и выше (In/Al), и ниже (Bi/C) ее, а для системы Cu/C [74] отсутствует вообще. Однако этот вывод нельзя считать окончательным, поскольку для системы Cu/C зависимость $\theta(T)$ исследована при небольших относительных переохлаждениях ($\Delta T/T_s \approx 0{,}12$), и, возможно, поэтому не обнаружена инверсия смачивания. Таким образом, имеющиеся к настоящему времени данные указывают на общность явления инверсии смачивания, однако еще не позволяют дать однозначный ответ на вопрос о ее механизме.

### 5.3. Размерный эффект при смачивании в переохлажденных конденсатах

Ранее (см. раздел 2.2), на примере ряда контактных систем (Sn/C, In/C, Bi/C, Pb/C, Au/C, Pb/Si) было показано, что смачивание аморфных нейтральных подложек жидкими металлами улучшается при уменьшении размеров микрокапель [17, 18]. Этот эффект является следствием уменьшения поверхностной энергии собственно капель металла $\sigma_l$ и межфазной энергии границы кап-
ля – подложка [17, 18] и исследован только для температур, пре-
вышающих температуру плавления металла. В то же время, из-
вестно, что кристаллизация малых частиц и, в частности, вакуум-
ных конденсатов происходит при значительных переохлаждениях
[77, 78], и для описания этого процесса необходимо знание как аб-
солютных значений краевых углов при соответствующих темпе-
ратурах, так и их размерной зависимости.

Такие исследования для островковых пленок олова на
аморфной углеродной подложке выполнены в работе [75]. Ре-
зультаты измерений краевых углов смачивания в системе Sn/C
при $T = 400$ К приведены на рис. 41, из которого видно, что для
переохлажденных капель, так же как и для равновесных [77,
78], наблюдается уменьшение краевого угла с уменьшением
размеров капель. Однако численные значения краевых углов
для капель одинаковых размеров оказываются другими, и зави-
симость $\theta(R)$ ($R$ – радиус капли) для переохлажденных капель
смещена в область меньших значений $\theta$ на величину $\Delta\theta \approx 15°$.

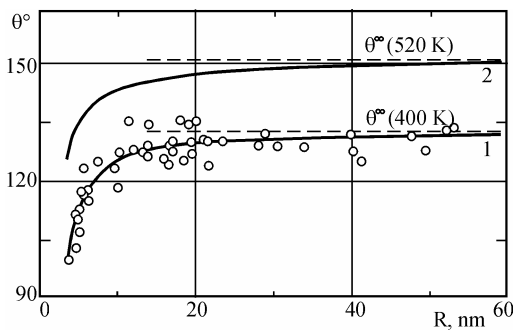

*Рис. 41. Зависимость угла смачивания от радиуса переохлажденных
(1. Т = 400 К [75]) и равновесных (2. Т = 520 К [17]) микрокапель олова*

Сравнение с известными результатами по размерному эф-
фекту смачивания при $T > T_s$ [17, 18] дает основания предпо-
лагать, что механизм эффекта для метастабильных капель (при

$T < T_s$) такой же, как и для равновесных, то есть обусловлен зависимостью от размера поверхностной энергии собственно капли и межфазной энергии границы капля – подложка [17, 18].

В рамках этого предположения обработка результатов в работе [75] проведена по методике, изложенной в разделе 2.2 с использованием соотношения (31), которое описывает равновесие микрокапли на твердой подложке.

Необходимо отметить, что численная обработка результатов эксперимента затруднена из-за отсутствия данных по поверхностной энергии металлов при больших переохлаждениях. Однако, согласно измерениям поверхностного натяжения ряда переохлажденных металлов [71], при переходе через точку плавления на зависимости $\sigma_l(T)$ не наблюдается каких-либо особенностей. Поэтому для определения $\sigma_l$ при $T < T_s$ в работе [75] использована линейная экстраполяция имеющихся литературных данных, относящихся к более высоким температурам [71]. Используя найденное таким образом значение $\sigma_l = 560$ мДж/м$^2$ при $T = 400$ K и предполагая, что характер размерной зависимости $\theta(R)$ сохраняется и в переохлажденном состоянии, то есть принимая для параметра $\alpha$ значение, соответствующее равновесным каплям ($\alpha = 0{,}25$ нм [17, 18]), из анализа экспериментальных данных с помощью выражения (31) в работе [75] были получены величины $\beta = 1{,}6$ нм, $\theta^{\infty} = 134°$ и $\sigma^{\infty}_{ul} = 500$ мДж/м$^2$. Для равновесных капель соответствующие параметры зависимости $\theta(R)$ оказываются несколько другими: $\beta = 1{,}0$ нм, $\theta^{\infty} = 152{,}7°$ и $\sigma^{\infty}_{ul} = 592$ мДж/м$^2$ [17, 18].

Таким образом, при исследовании размерного эффекта при смачивании в переохлажденных вакуумных конденсатах обнаружено, что зависимость $\theta(R)$ при $T > T_s$ подчиняется известным закономерностям, но величины параметров $\beta$, $\theta^{\infty}$ и $\sigma^{\infty}_{ul}$ выше и ниже $T_s$ оказываются различными.

# Заключение

Изложенные экспериментальные данные и их анализ показывают, что использование островковых вакуумных конденсатов для исследования смачивания позволило получить ряд оригинальных, общего характера результатов, имеющих принципиальное значение для физики поверхности и физико-химии поверхностных явлений. Эти результаты важны для теории и практики фазовых превращений, особенно для описания и управления процессом конденсации пересыщенного пара.

В первую очередь это относится к обнаружению, детальному исследованию и теоретическому описанию размерного эффекта при смачивании малыми жидкими частицами поверхности твердых тел. На этой основе впервые получены данные о размерной зависимости межфазной энергии границы раздела малая капля – твердая подложка. Важными представляются также данные по исследованию смачивания свободных пленок и развитые теоретические представления, позволившие определить поверхностную энергию свободных тонких пленок углерода.

Исследования в системах островковый конденсат – тонкая пленка убедительно показывают, что в таких системах, наряду с размерными эффектами, существенную роль играет также характер физико-химического взаимодействия компонентов.

Экспериментальные методы исследования смачивания, разработанные для изучения островковых конденсатов, найдут применение и при изучении других объектов.

В то же время, обнаруженная инверсия смачивания в островковых конденсатах, заключающаяся в переходе от несмачивания к смачиванию при понижении температуры переохлажденного состояния, нуждается в дальнейших исследованиях.

# Литература

# Содержание

Навчальне видання

Гладких Микола Тимофійович
Дукаров Сергій Валентинович
Фареник Володимир Іванович

**Змочування в острівцевих плівках**

Редактор Агаркова І.Ю.
Коректор Гавриленко О.В.